\documentclass[reprint, showpacs]{revtex4-1}
\usepackage{graphicx}
\usepackage{amsfonts}
\usepackage{natbib}
\usepackage{amsmath}
\usepackage{color}

\newcommand{\ep}{{\varepsilon}}
\newcommand{\be}{\begin{equation}}
\newcommand{\ee}{\end{equation}}
\newcommand{\bea}{\begin{eqnarray}}
\newcommand{\eea}{\end{eqnarray}}
\newcommand{\ba}{\begin{array}}
\newcommand{\ea}{\end{array}}
\newcommand{\lb}{\overline{l}}
\newcommand{\cL}{{\cal L}}
\newcommand{\cA}{{\cal A}}
\newcommand{\Vp}{{\rm Var}(p)}

\begin{document}
\title{Aspects of diffusion in the stadium billiard}

\author{\v Crt Lozej}
\author{Marko Robnik}

\affiliation{CAMTP - Center for Applied Mathematics and Theoretical
Physics, University of Maribor, Mladinska 3, SI-2000 Maribor, Slovenia, 
European Union}

\date{\today}

\begin{abstract}
We perform a detailed numerical study of diffusion in the 
epsilon-stadium of Bunimovich, and propose an empirical model of the 
local and global diffusion for various values of epsilon with the following
conclusions: (i) the diffusion is normal for all values of epsilon ($\le 0.3$) 
and all initial conditions, (ii) the diffusion constant is a parabolic function 
of the momentum  (i.e. we have inhomogeneous diffusion), (iii) the model 
describes the diffusion very well including the boundary effects, 
(iv) the approach to the asymptotic equilibrium steady state 
is exponential, (v) the so-called random model (Robnik et al 1997)
is confirmed to apply very well, (vi) the diffusion constant 
extracted from the distribution function in momentum space and the one 
derived from the second moment agree very well. The classical transport time,  
an important parameter in quantum chaos, is thus determined.   
\end{abstract}


\maketitle

\section{Introduction}

\begin{figure}
  \centering
  \includegraphics{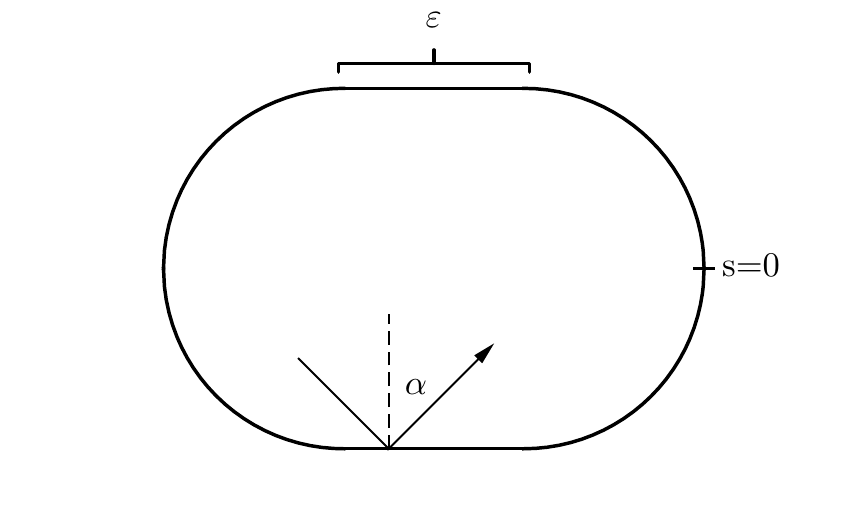}
  \caption{The geometry and notation of the stadium billiard of Bunimovich.}
  \label{figlr1}
\end{figure}

Billiard systems are very important model systems in classical and quantum chaos.
One of the most studied billiards is 
the stadium billiard introduced by L. Bunimovich \cite{Bun1979} in 1979, where it was proven to be
rigorously ergodic and mixing. It is also a K-system, as its maximal Lyapunov exponent is positive.
In Fig. \ref{figlr1} we show and define the geometry and our notation of the stadium.

The radius of the two half circles is unity, while the length of the straight line is $\ep$. By
$\alpha$ we denote the angle of incidence, which is equal to the angle of reflection at the collision
point.  The phase space is defined by the
Poincar\'e-Birkhoff coordinates $(s,p)$, where $s$ is the arclength parameter 
defined counterclockwise from $s=0$ to $\cL=2\pi+2\ep$ ($\cL$ is the length of the boundary), 
and the canonically conjugate momentum is $p=\sin \alpha$. As $s=0$ and $s=\cL$ are identified, we have a phase cylinder with the borders $p=\pm 1$.
We assume that the billiard particle has unit speed. 
The discrete bounce map of the billiard $\Phi$, connecting two successive collisions, 
$\Phi: (s,p) \rightarrow (s',p')$,  is area preserving (see e.g. \cite{Ber1981}). 
For the circle billiard $\ep=0$ the momentum $p$, 
which is also the angular momentum, is a conserved quantity, while for small $\ep>0$ we observe 
slow chaotic diffusion in the momentum space $p$. The maximal
Lyapunov exponent is positive for
all values of $\ep>0$. 
The first systematic study of the Lyapunov exponents in some representative chaotic
billiards, including the stadium billiard, was published by Benettin \cite{Ben1984}, who
has shown by
numerical calculations that for small $\ep$ the Lyapunov exponent goes as $\propto \sqrt{\ep}$.
The diffusion regime of slow spreading of an ensemble of initial conditions in the
momentum space has been observed in Ref. \cite{BCL1996}
 and  confirmed for $\ep \le 0.3$ in the present work. 
For larger $\ep > 0.3$ the diffusion regime is hardly observable, 
as the orbit of any initial conditions quickly spreads over the entire 
phase space, already after a few ten collisions.

The characteristic time scale on which transport phenomena occur in classical dynamical systems is termed the {\em classical transport time} $t_T$. This is the typical time that an ensamble of particles needs to explore the available phase space. In classical chaotic billiards the characteristic diffusion time in momentum space is the relevant estimate for $t_T$.
The present work was motivated by the study of chaotic billiards in the context of
quantum chaos\cite{Stoe,Haake}, in order to obtain good estimates of the characteristic
times $t_T$ which must be related/compared to the Heisenberg time $t_H$ 
for the purpose of assessing
the degree of quantum localization of chaotic eigenstates. The Heisenberg time
is an important time scale in any quantum system with a discrete energy spectrum,
defined as $t_H=(2\pi\hbar)/\Delta E$, where $\Delta E$ is the mean energy level
spacing, i.e. the mean density of states is $\rho(E)=1/\Delta E$. If $t_H/t_T$ is smaller
than $1$, we observe localization, while for values larger than $1$ we see
extended eigenstates and the {\em Principle of uniform semiclassical condensation} (PUSC) of Wigner functions applies. (See Refs. \cite{Rob1998,BR2010} and references therein.) 
Therefore a detailed investigation of the diffusion in the stadium billiard is necessary if
the localization of the Wigner functions or Poincar\'e-Husimi functions should be well
understood. 
This analysis is precisely along the lines of our recent works
\cite{BR2010,BR2013,BR2013A} for mixed type chaotic billiards, where
the regular and (localized) chaotic eigenstates have been separated and
the localization measure of the chaotic eigenstates has been introduced
and studied. It was shown that the spectral statistics is uniquely
determined by the degree of localization, which in turn is expected to
be a unique function of the parameter $t_H/t_T$. This kind of analysis 
has been performed also for the quantum kicked rotator 
\cite{Cas1979,Chi1981,Chi1988,Izr1990} by Chirikov, Casati,
Izrailev, Shepelyansky, Guarneri, and further developed by many others.
It was mainly Izrailev who has studied the relation between the
spectral fluctuation properties of the quasienergies (eigenphases)
of the quantum kicked rotator and the localization properties
\cite{Izr1988,Izr1989,Izr1990}. This picture has been recently extended
in \cite{BatManRob2013,ManRob2013,ManRob2014,ManRob2015}  
and is typical for chaotic time-periodic 
(Floquet) systems. Similar analysis in the case of the stadium as a time
independent system is in progress \cite{BLR2017}.

The study of diffusion in the stadium  goes back to the early works of 
Casati and coworkers \cite{BCL1996,CP1999}. An excellent review of 
classical and quantum chaotic billiards was published by Prosen \cite{Pro2000} 
with special interest in the quantum localization. Some of the analytic results
about the diffusion constant have been obtained by the study of an approximate
map \cite{BCL1996}, or of more general periodic Hamiltonian maps
\cite{DMP1989}, along with the special case of the sawtooth map \cite{CDMMP1990}.
 However, quite often
not only the pointwise orbits but even the statistical properties of 
conservative dynamical systems  exhibit extremely sensitive dependence
on the control parameters and on initial conditions, as exemplified
e.g. in the standard map by Meiss \cite{Mei1994}. 
The aforementioned studies used approximations of the stadium dynamics
in order to obtain analytical results.
Here we want to perform exact analysis of diffusion in the stadium billiard,
which unavoidably must rest upon the numerical  calculations
of the exact stadium dynamics.  Fortunately, the simple geometry of the stadium enables us to calculate the dynamics using analytical formulas subject only to round-off errors.

Thus far the study of momentum diffusion in the stadium billiard was, to the best of our knowledge, limited to the regime where ensembles are still narrow and far from the border of the phase space at $p=\pm 1$. The momentum diffusion there is normal and homogeneous and no border effects can be observed. In this work we extend this to include the global aspects of diffusion taking into account the finite phase space and the specifics of the billiard dynamics. As the stadium billiard is an archetype of systems with slow ergodicity many of our findings should be applicable to other systems sharing this trait.

The structure of the paper is  as follows. In section II we show that the 
chaotic diffusion is normal, but inhomogeneous, in section III we do a detailed 
analysis of the variance of the distribution function and 
explore its dependence on the shape parameter $\ep$ and the initial conditions, 
in section IV we examine the coarse grained dynamics and compare our 
results for the stadium with some preliminary results on a mixed type billiard defined in
\cite{Rob1983}, and in section V we discuss the results and conclude.

\section{Diffusion in the stadium billiard and the mathematical model}

The study of diffusion in the stadium billiard was initiated in \cite{BCL1996}, where
it was shown that for small $\ep$ we indeed see normal diffusion in the momentum space 
for initial conditions $p=0$ and uniformly distributed on $s$ along the boundary.
For sufficiently short times  (number of bounces), so that the spreading is close
to $p=0$, the diffusion constant can be considered as $p$-independent.  
Consequently the effects of the boundaries at $p=\pm 1$ are not yet visible. 
Moreover, it has been found \cite{CP1999} that the diffusion constant is 
indeed a function of the angular momentum, which for small $\ep$
coincides with $p$. Therefore we have to deal with inhomogeneous normal diffusion. 

Our goal is to elaborate on the details of this picture. 
We begin with  the diffusion
equation for the normalized probability density $\rho(p,t)$ in the $p$-space

\be \label{eq1}
\frac{\partial \rho}{\partial t} = \frac{\partial}{\partial p} 
\left( D(p) \frac{\partial \rho}{\partial p} \right).
\ee
The time $t$ here is the continuous time, related to the "discrete time"
$N$, the number of collisions, by $t=N \lb$, where $\lb$ is the average
distance between two collision points and the speed of the particle is unity.
In agreement with  \cite{CP1999} the diffusion constant is assumed in the form

\be \label{eq2}
D(p) = D_0(\ep) (1-p^2),
\ee
where $D_0(\ep)$ is globally an unknown, to be determined, function of the shape
parameter $\ep$. Note that in Ref. \cite{CP1999} the dependence of $D_0(\ep)$ on the
angular momentum was studied, while here we consider the dependence
on $p$.  It is only known that for sufficiently small $\ep$, 
smaller than a characteristic value $\ep_c \approx 0.1$ determined in the present work, 
or sufficiently larger than $\ep_c$, 
we have the power law  $D_0= \gamma \ep^{\beta}$, where the exponent $\beta$ is $5/2$ or $2$ correspondingly, while our $\gamma$ is a numerical prefactor, $\gamma \approx 0.13$ and $0.029$, respectively, 
also to be analyzed later on.
In the transition region, $\ep\approx \ep_c \approx  0.1$, we have no theoretical predictions
and also the numerical calculations are not known or well established so far.

As we see in Eq. (\ref{eq1}), the diffusion constant $D$ is defined in 
such a way that the probability current density $j$ is proportional to $D$ and 
the negative gradient of $\rho$, that is $j=-D\;\partial\rho/\partial p$.
Thus, the diffusion equation (\ref{eq1}) is just the continuity equation 
for the probability (or number of diffusing particles), 
as there are no sources or sinks.     
For $p\approx 0$ and at fixed $\ep$ we can regard $D$ as locally constant
$D\approx D_0$.  However, 
for larger $|p|$ we must take into account the dependence of $D$ on $p$.
Due to the symmetry the lowest correcting term in power expansion in $p$
is the quadratic one, with the proportionality  coefficient $\nu$. 
For larger $|p|$, close to $1$, the 
diffusion constant should vanish, so that near the border of the
phase space cylinder $p=\pm 1$ there is no diffusion at all. 
These arguments lead to the assumption
(\ref{eq2}), which will be a posteriori justified as correct in our detailed 
empirical model.

Let us first consider the case of locally constant $D$ at $p$ around $p_0$,
without the boundary conditions, i.e. the free diffusion on the real line $p$. 
Assuming initial conditions in the form of a Dirac delta distribution 
$\rho(p,t=0)=\delta(p-p_0)$ peaked at $p=p_0$, we recover the well known
Green function 

\be \label{eq3}
\rho(p,t) = \frac{1}{2\sqrt{\pi D t}} 
\exp\left( -\frac{(p-p_0)^2}{4D t}\right)
\ee
according to which the variance $\Vp$ is equal to

\be \label{eq4}
\langle (p-p_0)^2 \rangle = \int_{-\infty}^{\infty} \rho(p,t) (p-p_0)^2 dp 
= 2 D t.
\ee
This model is a good description for small $\ep$ and short times $t$.

However, for times comparable with the transport time
the boundary conditions must be taken into account.
There we assume that the {\em currents} on the boundaries $p=\pm 1$ 
must be zero, 
i.e. $\partial \rho/\partial p=0$, so that the total probability in the
momentum space is conserved and equal to unity. 
This will ultimately lead to the 
asymptotic equilibrium distribution $\rho(p,t) = 1/2$, with the variance
$\Vp=1/3$. The solution of the diffusion equation with these boundary conditions
reads \cite{Pol2002}

\bea \label{eq5}
\rho(p,t) = \frac{1}{2} + \sum_{m=1}^{\infty} A_m \cos (\frac{m\pi}{2}(p+1)) 
  \\ \nonumber
\times \;\exp\left( - \frac{D m^2\pi^2 t}{4}  \right).
\eea
Thus, the approach to the equilibrium $\rho=1/2$ is always exponential.  We will refer to this as the {\em homogeneous normal diffusion} model.

In the case of a delta function initial condition $\rho(p,t=0)=\delta(p-p_0)$, 
we find by a standard technique $A_m=\cos (\frac{m\pi}{2}(p_0+1))$.
In the special case $p_0=0$ we have

\be \label{eq6}
\rho(p,t) = \frac{1}{2}\sum_{-\infty}^{\infty} \cos (m\pi p) \exp(-m^2\pi^2Dt),
\ee
with the variance 

\be \label{eq7}
\Vp = \langle p^2 \rangle = \frac{1}{3} + 4 \sum_{m=1}^{\infty} 
\frac{(-1)^m}{m^2\pi^2} \exp(-m^2\pi^2D t),
\ee
which approaches exponentially the equilibrium value $\Vp=1/3$ at large time $t$

\be \label{eq8}
\Vp \approx \frac{1}{3} - \frac{4}{\pi^2} \exp( -\pi^2Dt),
\ee
where the higher exponential terms $m>1$ have been neglected. 

Next we want to understand the behavior of the diffusion when the full
general expression for the $p$-dependent diffusion constant $D$, defined 
in (\ref{eq2}), is taken into account. 
We find the solution (see also \cite{LL2007})
in terms of the Legendre polynomials  $P_l(p)$ as follows

\be \label{eq9}
\rho(p,t) = \sum_{l=0}^{\infty} A_l P_l(p) \exp (-l(l+1) D_0t),
\ee
where the expansion coefficients $A_l$ expressed  by the initial
conditions at time $t=t_0$ are

\be \label{eq10}
A_l = \frac{2l+1}{2} \int_{-1}^{1} P_l(p) \rho(p,t=t_0) dp.
\ee
It can be readily verified that the solution (\ref{eq9}) satisfies the
diffusion equation (\ref{eq1}) with $D=D_0(1-p^2)$ as in (\ref{eq2}). It
also satisfies the boundary conditions of vanishing currents at $p=\pm 1$, 
since  $D=0$ there.   Because the
set of all Legendre polynomials is a complete basis set of functions on the
interval $-1 \le p \le 1$, an arbitrary  initial condition may be
satisfied. Therefore (\ref{eq9}) is the general solution.
From (\ref{eq9}) we also see that $\rho(p,t)$
approaches its limiting value $A_0$ exponentially, and moreover, in (\ref{eq10}) 
that for any
normalized initial condition we have $A_0=1/2$. We will refer to this as the {\em inhomogeneous normal diffusion} model.

For a general diffusion constant $D(p)$ that is an even function of $p$, 
which in our case is due to the physical $p$-inversion 
symmetry in the phase space, we can derive a general equation for the
moments and variance of $p$. Starting from Eq. (\ref{eq1}) and using the
boundary conditions we first show that the total probability is conserved.
Second, for the centered initial condition $p_0=0$,  that is 
$\rho(p,t=0) = \delta(p)$, we find that the first moment vanishes 
$\langle p\rangle=0$, and for the time derivative of the variance we
obtain

\be \label{eq11}
\frac{d\langle p^2\rangle}{dt} = -4\rho(1,t)D(1)+2\int_{-1}^{1} \rho(p) 
\frac{d(pD(p))}{dp} dp.
\ee
In the special case $D=D_0(1 - \nu p^2)$ we get the differential equation

\be \label{eq12}
\frac{d\langle p^2\rangle}{dt} = -4\rho(1,t)D(1) +2D_0(1 - 3\nu  
\langle p^2\rangle).
\ee
This is an interesting quite general result. In our system we have
$\nu=1$, therefore $D(1) =0$, and we find for the variance the explicit
result by integration

\be \label{eq13}
\langle p^2\rangle = \frac{1}{3} \left(1 - \exp(-6D_0 t)\right).
\ee
Thus, again, the approach to equilibrium value $\Vp=1/3$ is exponential,
with the important {\em classical transport time} $t_T= 1/(6D_0)$.
If this equation is rewritten in terms of the discrete time $N$ 
(the number of collisions), then $t=N \lb$, and we find

\be \label{eq14}
\langle p^2\rangle = \frac{1}{3} \left(1 - \exp(-\frac{N}{N_T})\right),
\ee
where {\em the discrete classical transport time} $N_T$ is now
defined as

\be \label{eq15}
N_T = \frac{1}{6D_0\lb}.
\ee
Here $\lb$ is the average distance between two successive collision points.
We also define the discrete diffusion constant as $D_{\rm dis}=D_0 \lb$.
In the case of ergodic motion the mean free path $\lb$ 
as a function of the billiard area $\cA$ and the length $\cL$ is known to 
be \cite{Santalo}

\be \label{eq16}
\lb = \frac{\pi \cA}{\cL} =  \frac{\pi(\pi+2\ep)}{2\pi+2\ep} \approx \frac{\pi}{2}.
\ee
Thus, by measuring $D_{\rm dis}$ we determine $N_T$, which plays an important role
in quantum chaos when related to the Heisenberg time \cite{BR2013,BR2013A},
$t_H/t_T = 2k/N_T = 2\sqrt{E}/N_T$, as discussed in the introduction. 
Here $E=k^2$ is the energy of the billiard particle.

\begin{figure}
  \centering
  \includegraphics{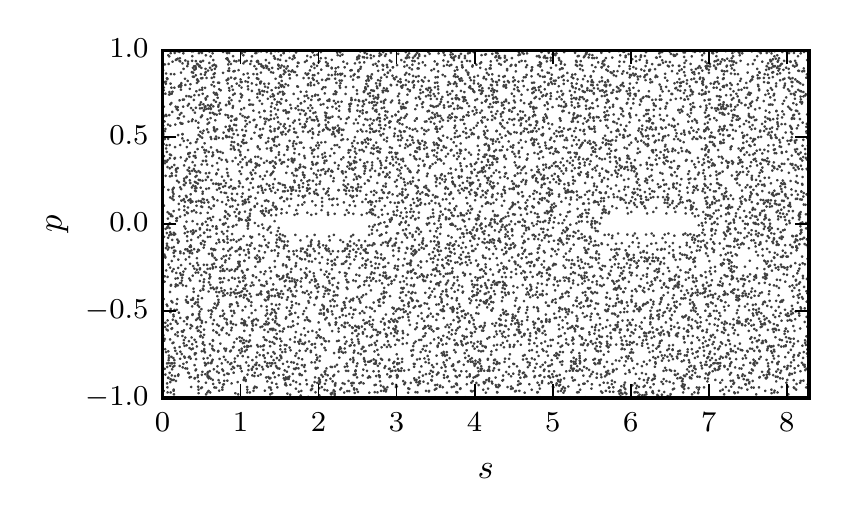}
  \caption{The phase space of the stadium for $\ep=1$, with $10^4$ bounces
along an orbit emanating from $(s=\pi/4,p=0)$,
showing the avoidance of the bouncing ball areas.}
  \label{figlr2}
\end{figure}

It is well known that the bouncing ball modes (the continuous family of
period two periodic orbits), within $s\in(\pi/2,\pi/2+\ep)$ and 
$s\in(3\pi/2+\ep,3\pi/2+2\ep)$  and $p=0$ present sticky objects in the 
classical phase space, as illustrated in Fig. \ref{figlr2}. If we choose
initial conditions inside these bouncing ball areas, we find a centrally
positioned delta peak which never decays. Moreover, even orbits close
to these bouncing ball areas stay inside for very long times,
because the transition times for exiting (and also entering) 
these regions are very large. Such correlations have been studied
in Refs. \cite{VCG1983,AHO2004}. 

\begin{figure}
  \centering
  \includegraphics{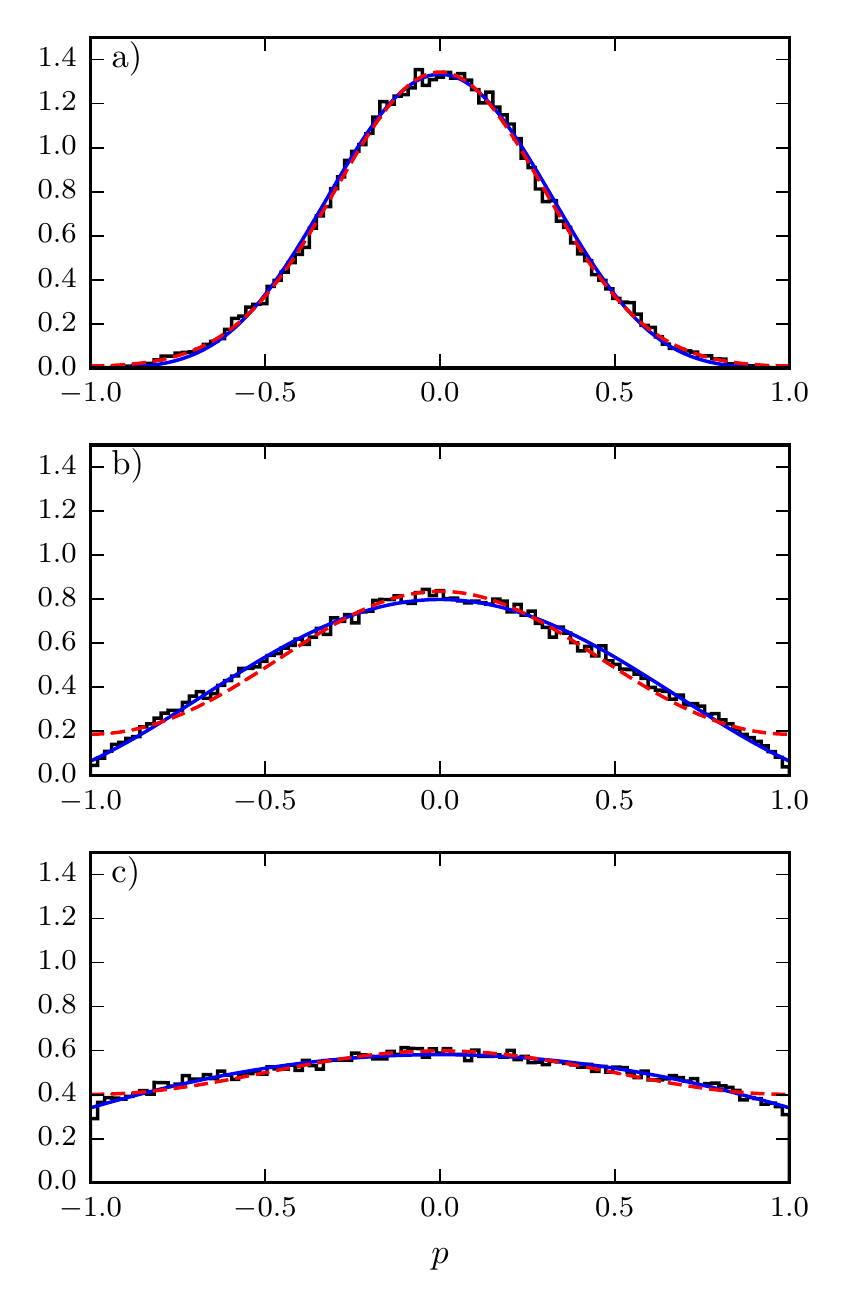}
  \caption{The distribution function $\rho(p,N)$ after $N=100$ collisions (a),
$N=300$ collisions (b) and $N=700$ collisions (c). The $10^5$ 
initial conditions at $p_0=0$ are as described in the text. $\ep=0.1$.
The blue full line is the
theoretical prediction for the inhomogeneous diffusion (\ref{eq9}), the
dashed red curve corresponds to the homogeneous diffusion (\ref{eq6}).
 }
  \label{figlr3}
\end{figure}

Therefore in studying the diffusion in the momentum space emanating from
$\rho(p,t=0)=\delta(p-p_0)$ we have used the initial conditions  $p_0=0$
and uniformly distributed over the $s$ excluding the two intervals
$s\in(\pi/2-\ep,\pi/2+2\ep)$ and $s\in(3\pi/2,3\pi/2+3\ep)$, to exclude the 
slowly decaying peak in the distribution located at $p=0$.
The result for $\ep=0.1$ is shown in Fig. \ref{figlr3}. As we see, the model of the
inhomogeneous diffusion Eq. (\ref{eq9}) is a significant improvement over
the model of homogeneous diffusion Eq. (\ref{eq6}) and works very well. 
The results are the same if we randomly vary the initial momenta according 
to a narrow uniform distribution $p_0\in [-0.01, 0.01]$.   

\begin{figure}
  \centering
  \includegraphics{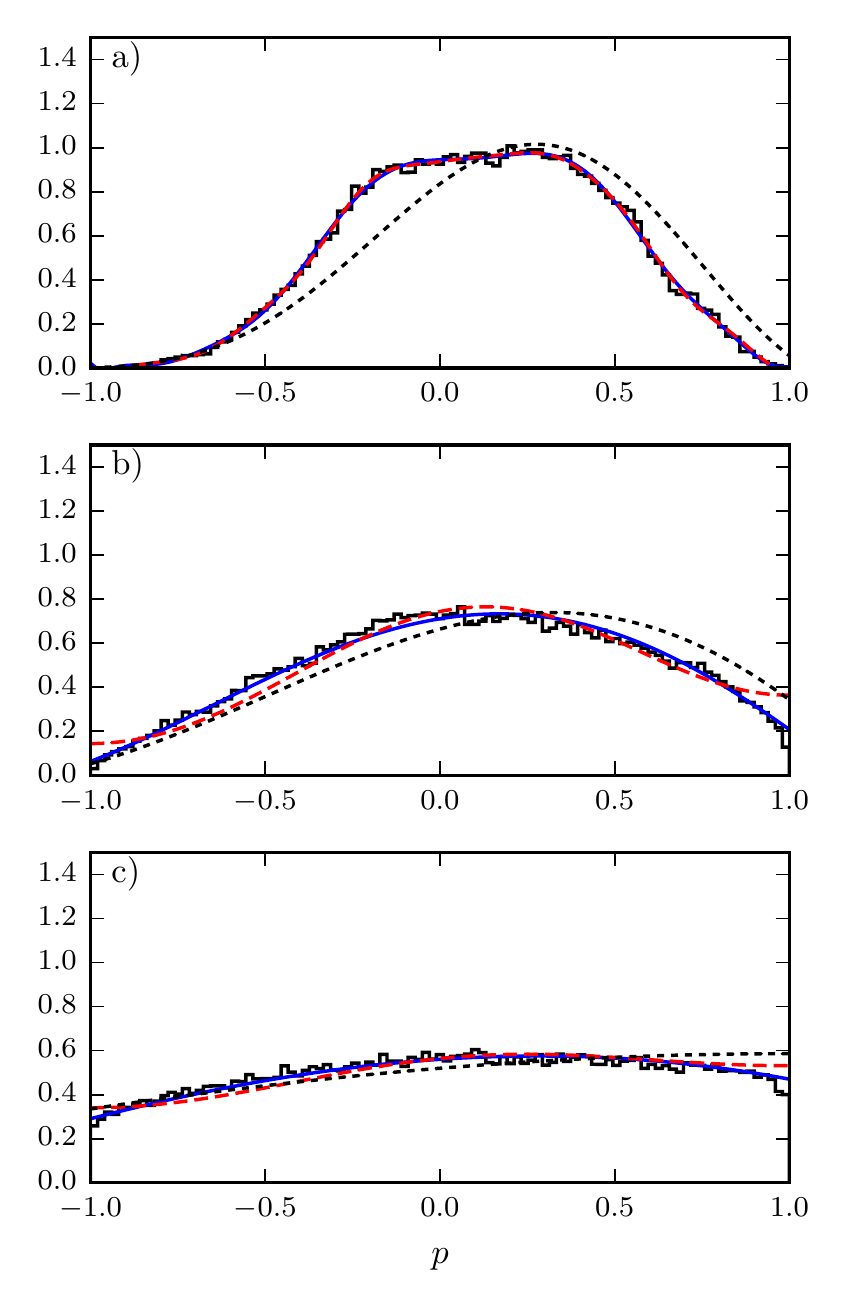}
  \caption{The distribution function $\rho(p,N)$ after $N=100$ collisions (a),
$N=300$ collisions (b) and $N=700$ collisions (c). The $10^5$ 
initial conditions at $p_0=0.25$ are uniformly distributed over $s\in [0,\cL)$,
and are taken at $N=100$ collisions. $\ep=0.1$. The blue full line is the
theoretical prediction for the inhomogeneous diffusion (\ref{eq9}), the
long-dashed red curve corresponds to the homogeneous diffusion (\ref{eq6}), while
the short-dashed black line is the theoretical prediction of
the inhomogeneous diffusion starting from the initial delta 
spike $\rho(p,t=0)=\delta(p-p_0)$ rather than from the delayed
histogram of (a).
 }
  \label{figlr4}
\end{figure}

\begin{figure}
  \centering
  \includegraphics{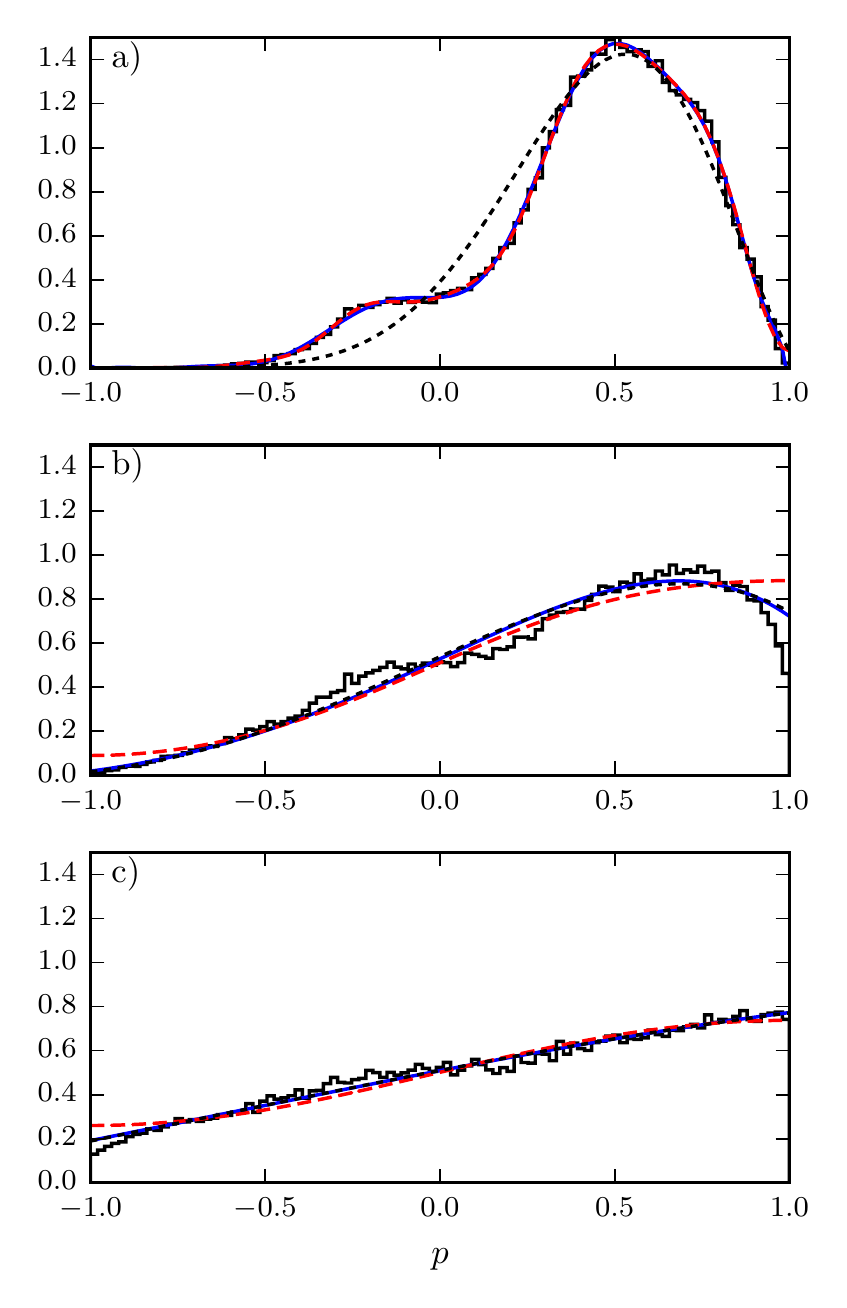}
  \caption{As in Fig.\ref{figlr4} but with $p_0=0.50$. 
 }
  \label{figlr5}
\end{figure}

\begin{figure}
  \centering
  \includegraphics{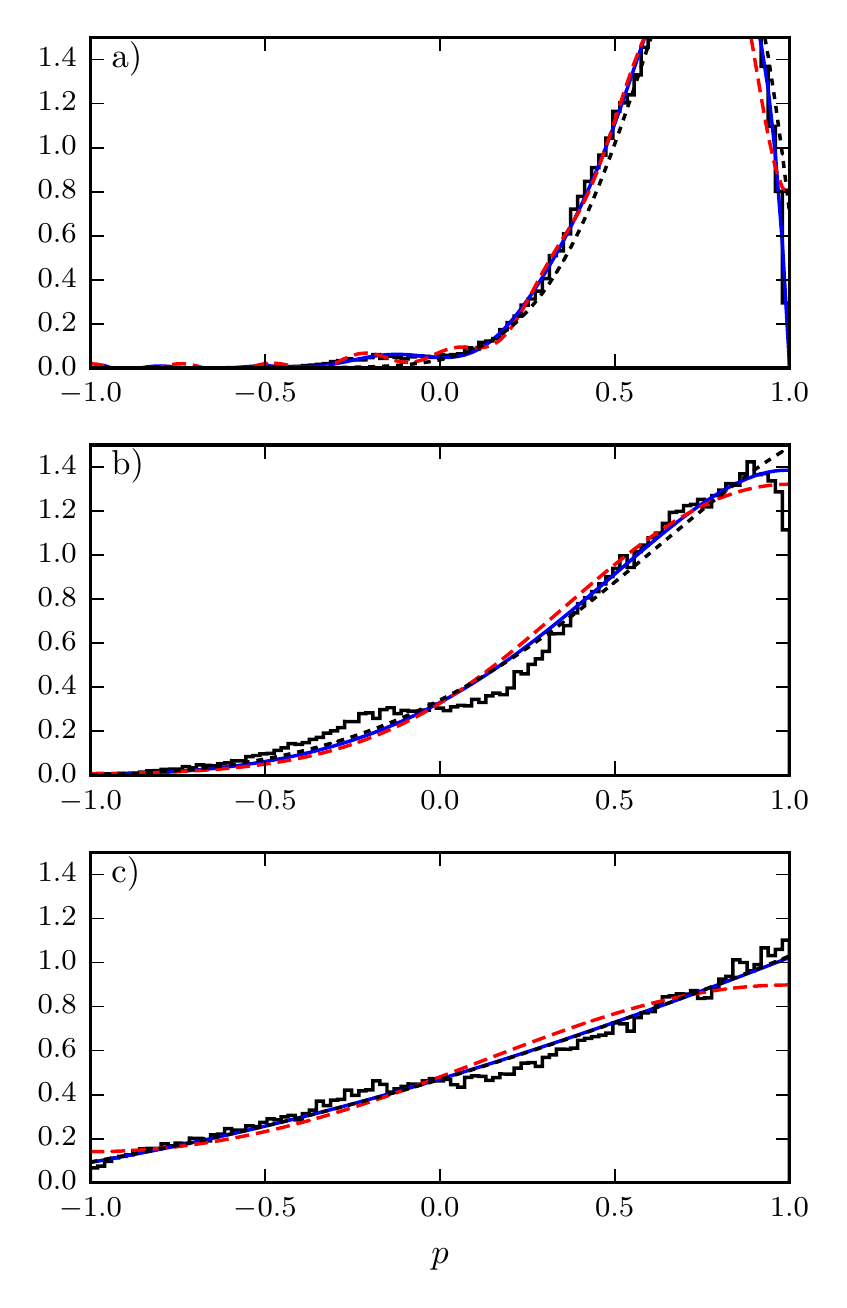}
  \caption{As in Fig.\ref{figlr4} but with $p_0=0.75$.
 }
  \label{figlr6}
\end{figure}

Although we shall study the dependence of statistical 
properties on initial conditions in
section III, we should explore the time evolution of the diffusion in the momentum space
for nonzero initial conditions $p_0\not=0$ already at this point,
starting from $\rho(p,t=0)=\delta(p-p_0)$, where
now the $10^5$ initial conditions are uniformly distributed over
all $s\in [0,\cL)$. It turns out that there is some transient time 
period, where the diffusive regime is not yet well established, 
which we demonstrate for $\ep=0.1$ in Figs. \ref{figlr4} - \ref{figlr6}
for $p_0=0.25, 0.50$ and $0.75$, correspondingly. The black short-dashed
curve corresponds to the theoretical prediction based on the
inhomogeneous diffusion model (\ref{eq9}) starting with the
initial delta spike $\rho(p,t=0)=\delta(p-p_0)$, 
and the initial conditions are uniform on all $s\in[0,\cL)$. 
The red long-dashed
curve corresponds to the homogeneous diffusion model (\ref{eq6}), 
the blue full line corresponds to the inhomogeneous diffusion
model (\ref{eq9}). In both latter cases the initial conditions were
taken from the histogram at the time of $100$ collisions,
and the coefficients  in Eqs. (\ref{eq5},\ref{eq6},\ref{eq9},\ref{eq10})
were determined.
The delay of $100$ collisions has been chosen due to the initial
transient behaviour where the diffusion is not yet well defined. 
Nevertheless, the time evolution of the diffusion
 excellently obeys the inhomogeneous law (\ref{eq9})
for longer times. 

In these plots we observe qualitatively good agreement with the theory,
except for some visible deviation of the numerical histogram from the
theoretical prediction, around $p\approx 0$ in Figs. 
\ref{figlr5}-\ref{figlr6}. We believe that these effects
are related to sticky objects around the 
bouncing ball areas whose existence is demonstrated in the phase 
space plot of Fig. \ref{figlr2}, and is thus a system-specific feature,
which would disappear in a "uniformly ergodic" system. However, 
it must be admitted that chaotic billiards often have such 
continuous families of marginally stable orbits 
\cite{Alt2008}.

\section{Analysis of the variance of the distribution function}

In order to determine the value of the diffusion constant $D_0$ and its
dependence on $\ep$ as defined in Eqs. (\ref{eq1}-\ref{eq2}) we can use
either the evolution of the entire distribution function $\rho(p,t)$
or of its second moment, the variance. When this was done in special
cases, agreement has been found. However, the second moment of distribution
function is much more stable than the distribution function itself, so
we have finally decided to use the variance of $p$ to extract the value
of $D_0$ from Eq. (\ref{eq13}), where the delta spike initial
condition at $p_0=0$ is assumed. This approach would be ideal, if our 
model  of the inhomogeneous diffusion in Eqs. 
(\ref{eq1},\ref{eq2},\ref{eq9},\ref{eq10}) were exact.
 
However, this is not the case due to the bouncing ball regions
described in the previous section and the fact that the diffusive 
regime is not yet well established for short times. Moreover, even if the model were exact,
Eq. (\ref{eq13}) does not apply to initial conditions at nonzero $p_0$. 
The variance for the more general initial conditions $\rho(p,t=0)=\delta(p-p_0)$ 
is easily obtained by calculating the first two moments of the distribution (\ref{eq9}). 
This is done by inserting the initial conditions into Eq. (\ref{eq10}) and using 
the orthogonality relations of the Legendre polynomials. The average momentum is given by
\begin{equation}\label{eq17a}
\langle p\rangle =p_0\exp(-2D_0 t),
\end{equation}
and the variance by
\begin{multline}\label{eq17}
\Vp = \langle p^2\rangle - (\langle p\rangle)^2 =\\
 = \frac{1}{3} \left(1 -  \exp(-6D_0 t)\right)+{p_0}^2\exp(-6D_0 t)-\\-{p_0}^2\exp(-4D_0 t).
\end{multline}
As we see there is no clear way to define the transport time since two exponential functions with different exponents are present. Furthermore, as we saw earlier in Figs. \ref{figlr4}-\ref{figlr6} the diffusive regime is only well established after enough time has passed. We therefore chose to empirically generalize equation Eq.(\ref{eq14}) by the introduction of a prefactor $C$ as follows  
\be \label{eq18}
\Vp = \frac{1}{3} \left(1 - C \exp(-\frac{N}{N_T})\right).
\ee
As we shall see the prefactor $C$ effectively compensates the generalized initial conditions and allows us to estimate the transport time.

\begin{figure}
  \centering
  \includegraphics{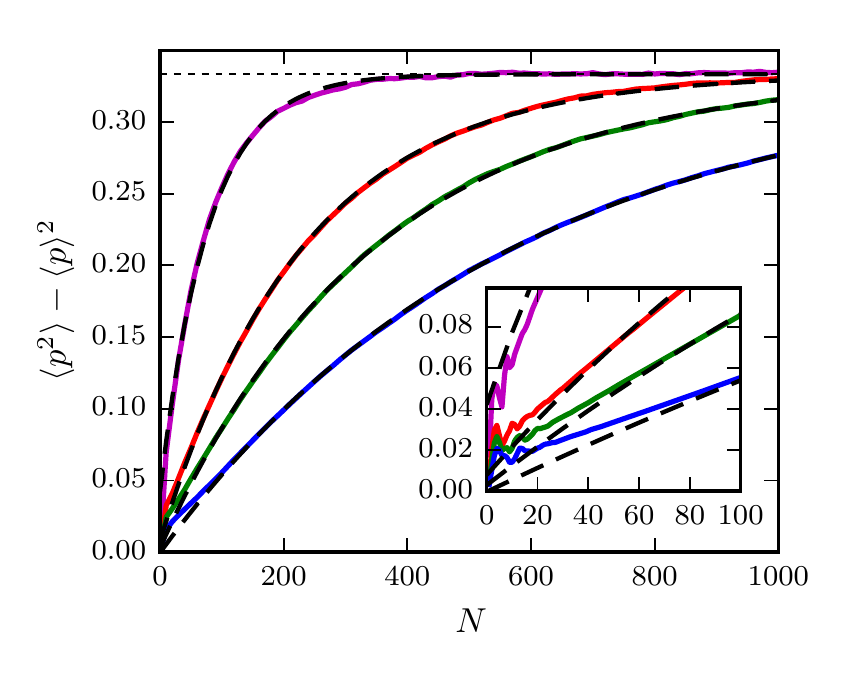}
  \caption{The variance of $p$ as a function of the number
of collisions $N$ for four different values of $\ep$, starting 
at the initial condition $p_0=0$. From top to bottom we show the numerical results for $\ep=0.2$ (magenta), $\ep=0.12$ (red), $\ep=0.1$ (green), $\ep=0.08$ (blue) calculated with $10^5$ initial conditions.
The theoretical fitting curves according to Eq. (\ref{eq18}) are shown with black dashed lines. The inset is a magnification of the first 100 bounces and shows the initial non-diffusive phase.
 }
  \label{figlr9}
\end{figure}

In Fig. \ref{figlr9} we show the evolution of the variance
as a function of the number of collisions, for four different
values of $\ep=0.08, 0.10, 0.12, 0.20$. The initial conditions
are the same as in Fig. \ref{figlr3} of Sec. II.
The agreement with the empirical model (\ref{eq18}) (which in this case coincides with the theoretical prediction (\ref{eq14}) if $C=1$) is excellent. The inset shows the initial non-diffusive phase of the dynamics.
The fitting procedure was as follows: first, $N_T$ has been extracted from the best fitting of all data, and then the fitting was repeated by excluding
the first $20\%$ of collisions, but not less than $20$ of them, 
up to maximum of $300$ collisions, in order to be in the optimal
interval for the determination of the two fitting parameters
$N_T$ and $C$ and to exclude the non-diffusive phase.

\begin{figure}
  \centering
  \includegraphics{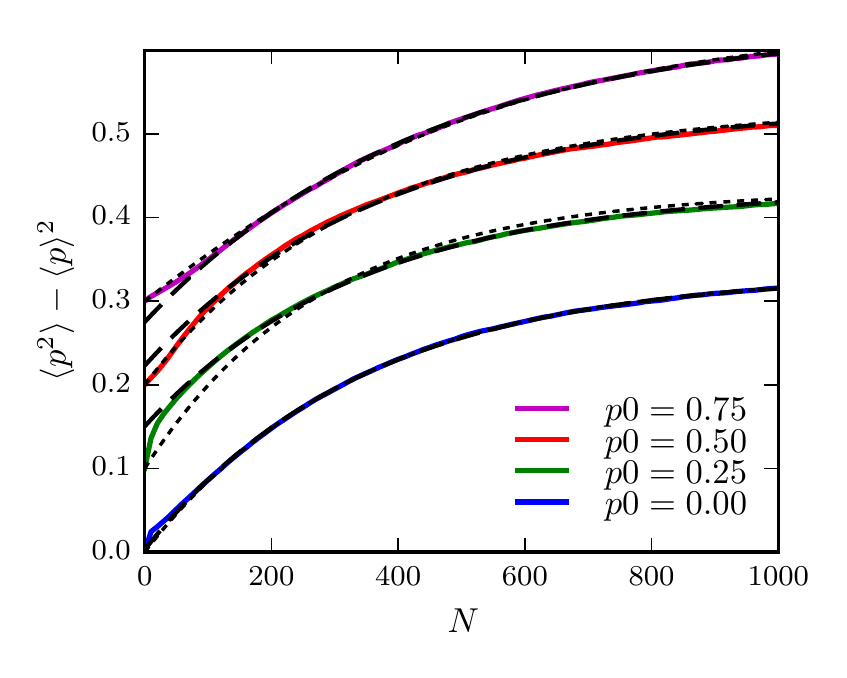}
  \caption{The variance as a function of the number
of collisions $N$ for $\ep=0.1$, starting 
at four initial conditions $p_0=0, 0.25, 0.50, 0.75$.
Each curve is shifted upwards by $0.1$ in order to avoid the overlapping 
of curves. They all converge to $1/3$ when $N\rightarrow \infty$. The dotted black curve shows the theoretical prediction for the variance (\ref{eq17}), while the dashed curve shows the empirical model (\ref{eq18}). 
 }
  \label{figlr10}
\end{figure}

Finally, we take a fixed value of $\ep=0.1$ and observe the variance
as a function of discrete time $N$ for various values of the initial 
condition $\delta(p-p_0)$, $p_0=0., 0.25, 0.50, 0.75$, and use the same 
fitting procedure. The initial conditions are the same as in Figs. \ref{figlr3}
- \ref{figlr6}, correspondingly.
Again, the agreement with the empirical 
formula (\ref{eq18}) is excellent, as is seen in Fig. \ref{figlr10}.
The variance is well described also by the theoretical prediction Eq. (\ref{eq17}), particularly for larger values of $p_0$. From there we may extract the value of $D_{\rm dis}=D_0 \lb$. Note that for $p_0=0$ the two descriptions coincide. Alternatively we could also extract the values of $D_{\rm dis}$ from the average of the momenta using Eq. (\ref{eq17a}). This yields equivalent results for values of $p_0>0.35$, but for lower nonzero values, Eq. (\ref{eq17a}) fails to correctly describe the numerical time dependent averages. This is because the initial non-diffusive phase significantly changes the average of the distribution of momenta from the one predicted from the initial delta distribution (compare the histogram with the black dashed line in Fig. \ref{figlr4}). This effect is diminished for $p_0>0.35$, probably because the peak of the distribution is further from area of phase space near the marginally unstable bouncing ball orbits (see Figs. \ref{figlr5}
- \ref{figlr6} ).

In table 1 we present the list of values of $N_T$ at $p_0=0$, as a function of 
$\ep$, extracted by the described methodology. They are important
in understanding the quantum localization of chaotic eigenstates as
previously discussed.

\begin{table}
\center
    \begin{tabular}{ | p{1.7cm} | p{1.7cm}  || p{1.7cm} | p{1.7cm}|}
    \hline
    \multicolumn{4}{|c|}{Transport times} \\
    \hline
    $\ep$ & $N_T$ & $\ep$ & $N_T$\\ \hline
    0.001 & $3 \times 10^{7}$ & 0.105 & 303\\ \hline
    0.005 & $4.3 \times 10^{5}$& 0.110 & 275\\ \hline
    0.010 & $7.8 \times 10^{4}$& 0.115 & 253\\ \hline
    0.015 & $2.9\times 10^4$ & 0.120 & 233\\ \hline
    0.020 & $1.4 \times 10^4$& 0.125 & 215\\ \hline
    0.025 & 8410 & 0.130 & 299\\ \hline
    0.030 & 5520 & 0.135 & 186\\ \hline
    0.035 & 3750 & 0.140 & 172\\ \hline
    0.040 & 2760 & 0.145 & 161\\ \hline
    0.045 & 2110 & 0.150 & 150\\ \hline
    0.050 & 1630 & 0.155 & 141\\ \hline
    0.055 & 1340 & 0.160 & 131\\ \hline
    0.060 & 1100 & 0.165 & 123\\ \hline
    0.065 & 907 & 0.170 & 115\\ \hline
    0.070 & 767 & 0.175 & 108\\ \hline
    0.075 & 647 & 0.180 & 102\\ \hline
    0.080 & 560 & 0.185 & 95\\ \hline
    0.085 & 494 & 0.190 & 90\\ \hline
    0.090 & 433 & 0.195 & 86\\ \hline
    0.095 & 386 & 0.200 & 82\\ \hline
    0.100 & 341 &   &  \\ \hline
    \end{tabular}\\
    \caption{The discrete transport time $N_T$ (number of collisions) 
as function of $\ep$ as defined in Eq. (\ref{eq18}), with initial conditions at $p_0=0$, and therefore Eq. (\ref{eq14}) also applies. The sizes of the ensembles were $10^5$ for $\ep<0.02$, $2.5\times10^5$ for $0.02\leq \ep<0.08$ and $10^6$ for $\ep>0.08$.}
\end{table}

\begin{figure}
  \centering
  \includegraphics{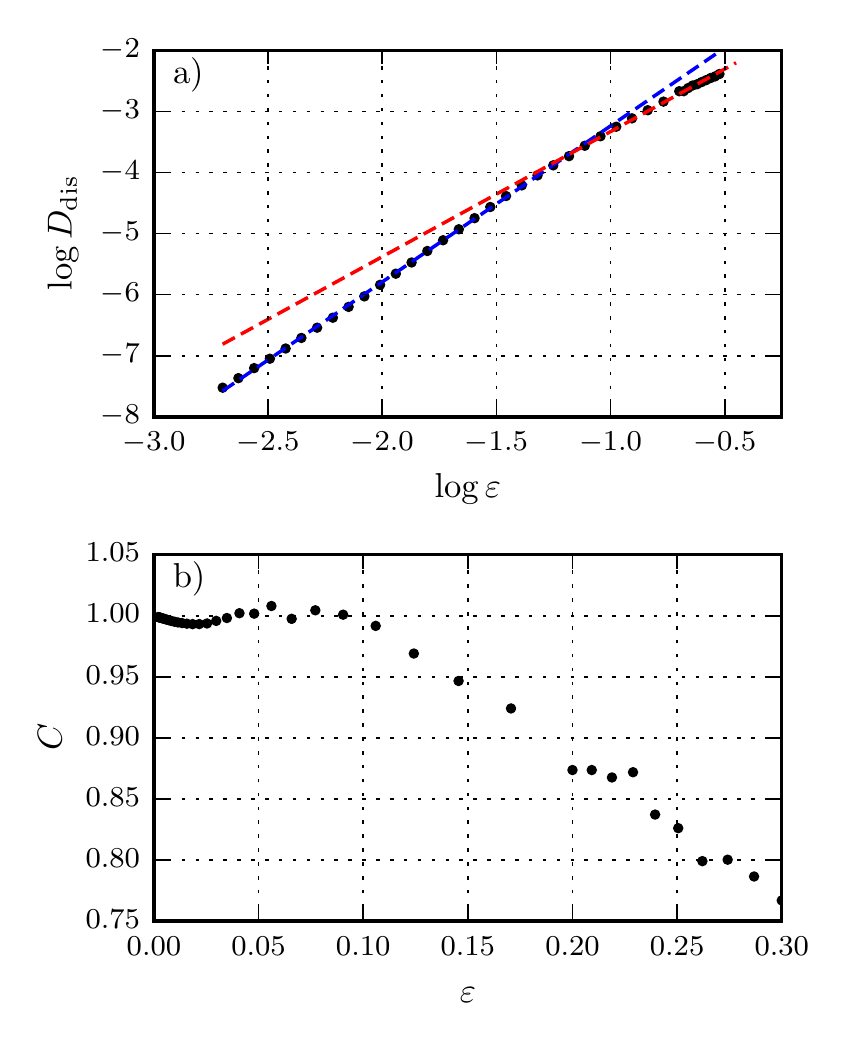}
  \caption{$D_{\rm dis}= D_0\lb=1/(6N_T)$ and $C$ as 
functions of $\ep$, at $p_0=0$,
using the Eq. (\ref{eq18}). The $\log$ is decadic. 
Ideally, $C$ should be unity according to Eqs. (\ref{eq13},\ref{eq14}).
 }
  \label{figlr7}
\end{figure}

In Fig. \ref{figlr7} we show the result for the
diffusion constant in terms of the discrete time $N$,
that is $D_{\rm dis}= D_0\lb$, as a function of $\ep$, as well
as $C$ as a function of $\ep$, with the initial conditions
$\rho(p,t=0)=\delta(p-p_0)$ at $p_0=0$.

We clearly observe the confirmation of the two limiting power laws
$D_{\rm dis} \propto \ep^{5/2}$ for small $\ep$ and 
$D_{\rm dis} \propto \ep^2$ for large $\ep$. In the transition
region $\ep\approx \ep_c \approx 0.1$ which is about half
of a decade wide, the analytic description is unknown.

\begin{figure}
  \centering
  \includegraphics{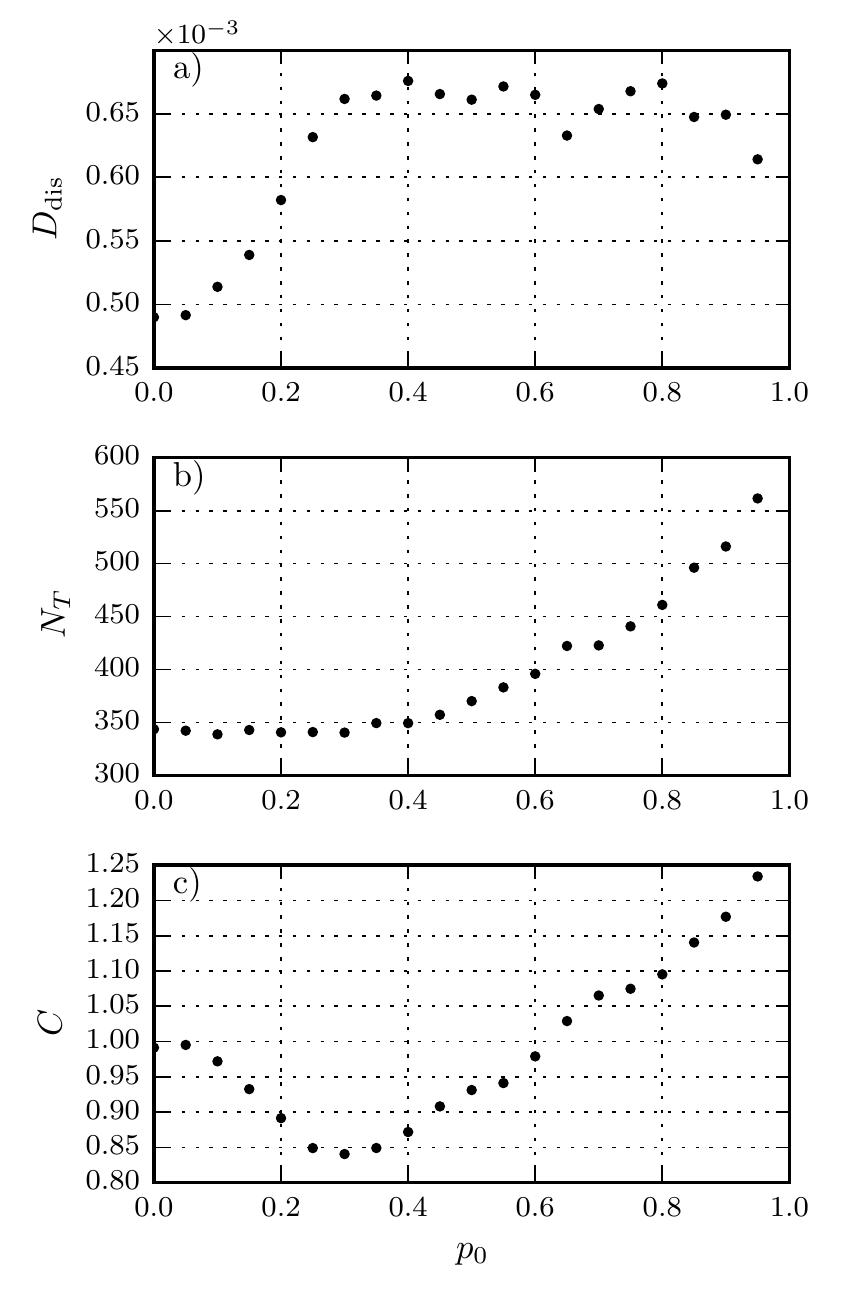}
  \caption{$D_{\rm dis}= D_0\lb$ from (\ref{eq17}), and $N_T$ with $C$ from (\ref{eq18}) as functions of $p_0$ for $\ep=0.1$.
 }
  \label{figlr8}
\end{figure}

The dependence of $D_{\rm dis}$ in accordance with Eq. (\ref{eq17}) on the parameter $p_0$ for the special case $\ep=0.1$ is shown in Fig. \ref{figlr8} (a). The value of $D_{\rm dis}$ is minimal at $p_0=0$. This may be because the phase space contains sticky objects near the marginally unstable bouncing ball orbits. Here we must understand that at larger times, asymptotically, the initial conditions are forgotten, and we expect that $D_{\rm dis}$ tends to a constant value which is the case. The values of $D_{\rm dis}$ at $p_0=0.75$ exhibit the same power law dependences on $\ep$ as those at $p_0=0$.  $N_T$ and $C$ in accordance with Eq. (\ref{eq18}) are shown in Fig. \ref{figlr8} (b-c). The value of $N_T$ increases for larger values of $p_0$.
This is an effect of the local transport being slower in the vicinity of the border $p=\pm 1$ due to the parabolic diffusion law (\ref{eq2}). The shortest estimate for the classical transport time, the one at $p_0=0$, is the one relevant for the study of localization of the eigenstates of the quantum billiard.

\section{Comparison of chaotic diffusion in a mixed type billiard}

In this section we study the coarse grained dynamics of the stadium billiard. We partition the phase space into a grid of cells and record the number of times each cell is visited by the orbit.
It is interesting to briefly discuss the observed differences between  
ergodic systems like the stadium and the behavior in the chaotic components of
mixed type systems. Examples are the billiard introduced in \cite{Rob1983}, the border of which is given by a conformal mapping of the unit circle in the complex plane $|z|=1$
\begin{equation}
z \rightarrow z+\lambda z^2,
\end{equation}
at various shape parameter values $\lambda$ 
and the standard map \cite{Mei1994}. 
In the case of $\lambda=1/2$ the billiard was proven to be ergodic \cite{Mar1993}. Because this billiard is strongly chaotic the classical transport time is of the order of a few 10 bounces. An analysis of the diffusion along the lines of the previous sections is therefore not possible.

\begin{figure}
  \centering
  \includegraphics{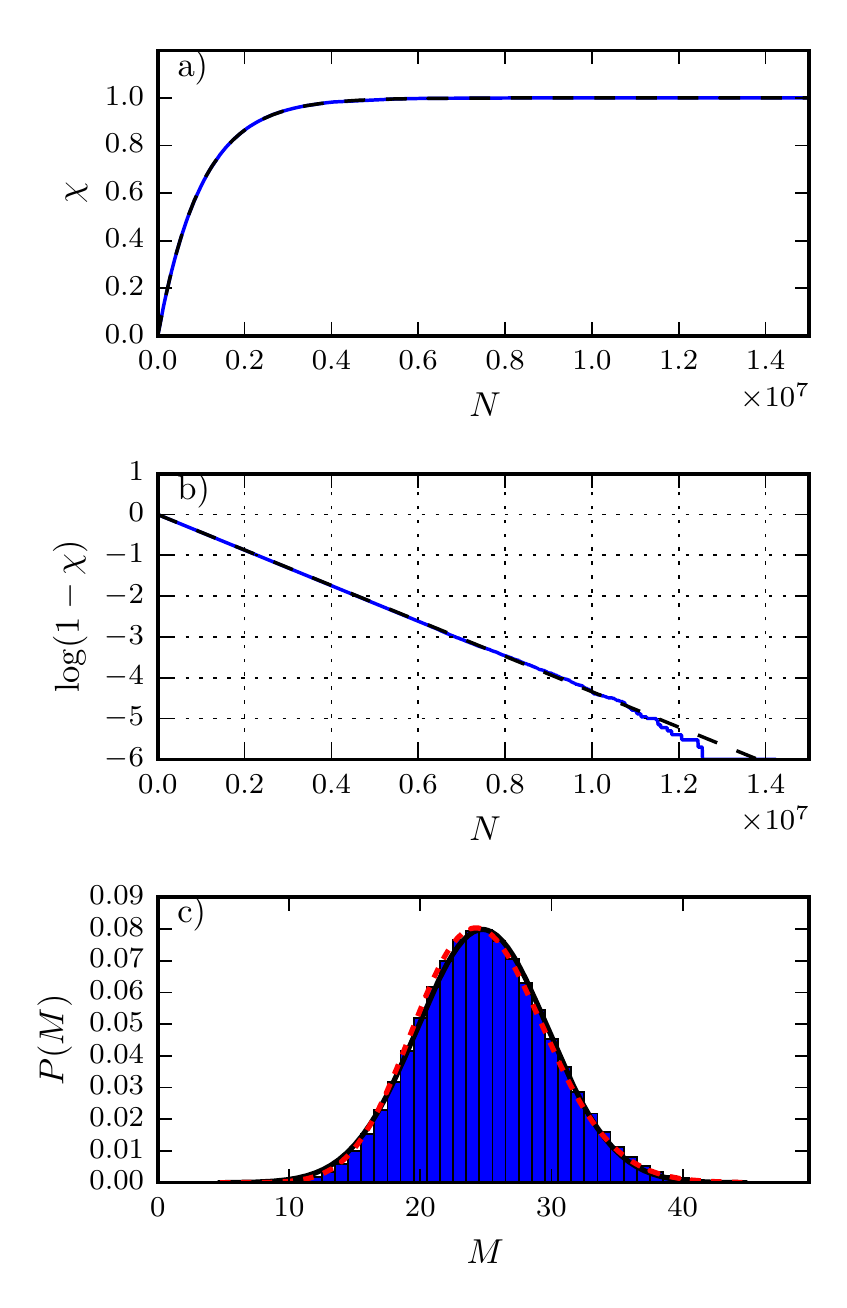}
  \caption{The filling of the cells $\chi(N)$ for the stadium billiard $\ep=0.1$ in the lin-lin plot (a), 
the log-lin plot (b) and the distribution of cells with the occupancy number $M$ (c). 
The black  dashed curve in (a) and (b) is the random model Eq. (\ref{eq20}).
There are $25\times 10^6$ collisions and $N_c=10^6$ cells, so that the mean occupancy number $\mu =\langle M \rangle = 25$.  The chaotic orbit has the initial conditions $(s=\pi/4, p=0)$. The black full 
curve is the best fitting Gaussian, while the red dashed curve is the best fitting Poissonian distribution. The theoretical values $\mu$ and $\sigma^2= \mu b =\mu (1-a)$ and their numerical values agree very well.}
  \label{figlr11}
\end{figure}

\begin{figure}
  \centering
  \includegraphics{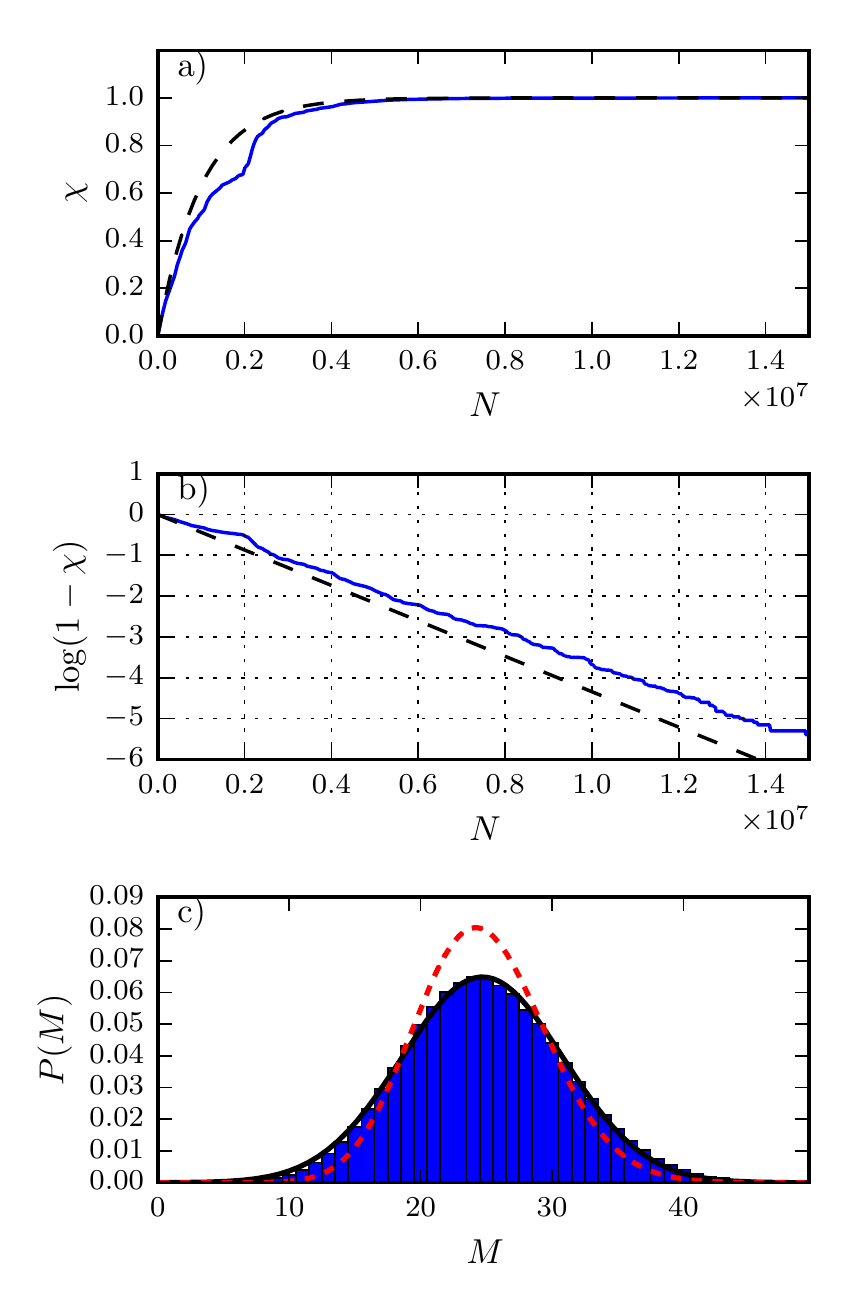}
  \caption{As in Fig. \ref{figlr11} but $\ep=0.01$. Some deviations from the 
random model are seen, which are always negative due to the sticky objects in 
the phase space which delay the diffusion.}
  \label{figlr12}
\end{figure}

The first important observation is, that
the so-called {\em random model} (Poissonian filling of the coarse grained 
network of cells in the phase space) introduced in \cite{Rob1997}, works very
well in the stadium and also in other ergodic systems like $\lambda=1/2$ billiard.
In the process of filling, the cells are considered as filled (occupied) as soon as the orbit visits them.
The approach to the asymptotic value $1$ for the relative size of
the filled chaotic component $\chi$ as a function of the discrete time 
(number of collisions) $N$ is exponential,

\be \label{eq20}
\chi(N) = 1 -\exp \left( -\frac{N}{N_c} \right).
\ee
where $N_c$ is the number of cells.
This is demonstrated in Fig.\ref{figlr11} for the case $\ep=0.1$, where
chaos, dependence on initial conditions due to the large Lyapunov exponent, 
is strong, and agreement with the random model is excellent, 
while for the case $\ep=0.01$ chaos is weak, and the agreement is not so
good as seen in Fig. \ref{figlr12}.

In both cases we show the lin-lin plot
of (\ref{eq20}) in (a), and also the log-lin plot in (b) for $1-\chi$. In
(c) we show the distribution of the occupancy number $M$ of
cells, which clearly is very close to a Gaussian. The size of the grid of
cells is $L=1000$, thus $N_c=L^2= 10^6$. 

The latter observation can be easily explained by the following theoretical
argument within the Poissonian picture. We start an orbit in one of the 
$N_c$ cells of the chaotic region and follow its evolution for a fixed 
number of collisions $N$. Let $a=1/N_c$ be the uniform 
probability that at the given discrete time $N$ one of the cells will 
be visited (by the orbit), while its complement $b=1-a$ 
is the probability that the cell will not be visited. As we assume 
absence of any correlations between the visits, the calculation of
the distribution of the occupancy $M$ of the cells is easy: The probability
$P(M)$ to have a cell containing $M$ visits is simply the binomial
distribution

\be \label{eq21}
P_B(M) = {{N}\choose{M}} a^M b^{N-M},
\ee
which has the exact values for the mean and variance

\be \label{eq22}
\mu= \langle M \rangle =Na, \;\;\; \sigma^2 = \langle (M-\mu)^2 \rangle=Nab 
= \mu b.
\ee
For sufficiently large $N$ this can be approximated by the Gaussian with the
same $\mu$ and $\sigma^2$,

\be \label{eq23}
P_G(M)= \frac{1}{\sqrt{2\pi\sigma^2} }  
\exp\left( -\frac{(M-\mu)^2}{2\sigma^2}  \right).
\ee
In the Poissonian limit $a\rightarrow 0$ and $N\rightarrow \infty$, but
$\mu=aN={\rm const.}$, we find the Poissonian distribution

\be \label{eq24}
P_P(M)= \frac{\mu ^M e^{-\mu}}{M!}.  
\ee
We see that the mean value and the variance agree with the theoretical
prediction $\mu=N/N_c=25$ and $\sigma^2=\mu b\approx \mu=25$, so that
the standard deviation $\sigma=5$, in the case of $\ep=0.1$. 
However, in the case $\ep=0.01$ we see a quantitative discrepancy with
the random model prediction in Fig. \ref{figlr12} (a), but nevertheless
the approach to the asymptotic values is exponential with a slightly
different coefficient (b). Due to the sticky objects the filling of the cells 
is slower.  The variance of the distribution in (c) 
is $38$,  which is larger than the predicted value $\mu b=25$,
meaning that the relative fraction of more and of less richly
occupied cells is larger than expected by the binomial distribution. 
Note that the Poisson distribution with the same $\mu$ (dashed red) 
significantly deviates from the histogram.

\begin{figure}
  \centering
  \includegraphics{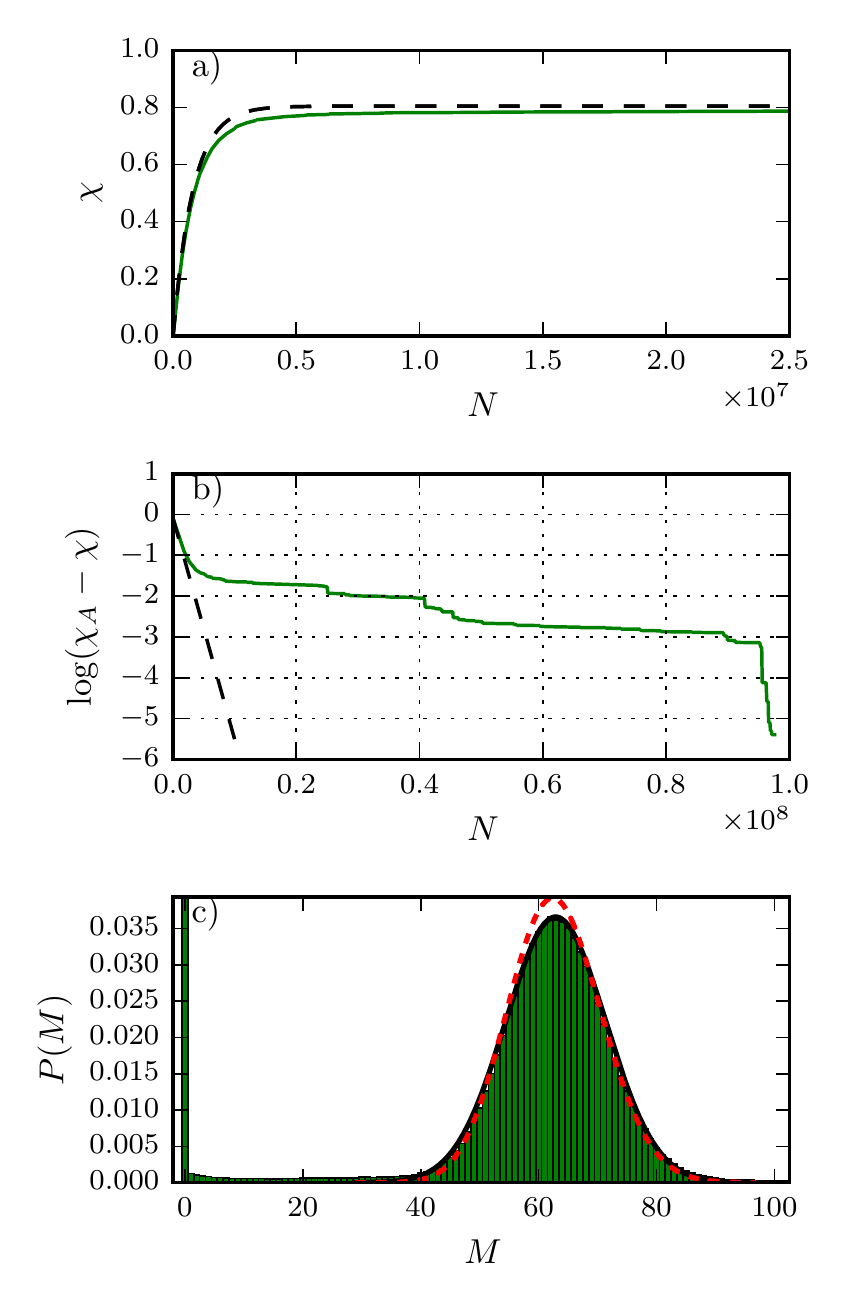}
  \caption{As in Fig. \ref{figlr11} but for the mixed type billiard
\cite{Rob1983} with $\lambda=0.15$, for an orbit with $10^8$ collisions. 
The asymptotic value of $\chi$ is $\chi_A=0.80$, as we see in (a). 
Large deviations from the 
random model are seen also in (b), which are always negative due to the sticky objects in 
the phase space which delay the diffusion. In (c) we see the distribution
of cells at time $N=5\times 10^7$ according to the occupancy $M$, which now consists of the
the Gaussian bulge corresponding to the cells of the largest chaotic region,
and the delta peak at $M=0$ corresponding to the regular and other
smaller chaotic regions.
The initial condition (using the standard Poincar\'e-Birkhoff coordinates) 
is $(s=4.0,p=0.75)$. 
}
  \label{figlr13}
\end{figure}

In the chaotic components of mixed type systems things are different.
The random model does not work well, the approach to the equilibrium value 
is not exponential, but instead is perhaps a power law 
as reported by Meiss \cite{Mei1994},  or even something else
as observed in our work \cite{LR2017}.
Here we just show for comparison in Fig. \ref{figlr13} the time 
dependence of the relative fraction of occupied cells $\chi(N)$
for the billiard introduced in \cite{Rob1983} with the shape parameter
$\lambda=0.15$ (a slightly deformed circle), in analogy with Figs. \ref{figlr11}-\ref{figlr12}.

\begin{figure}
  \centering
  \includegraphics{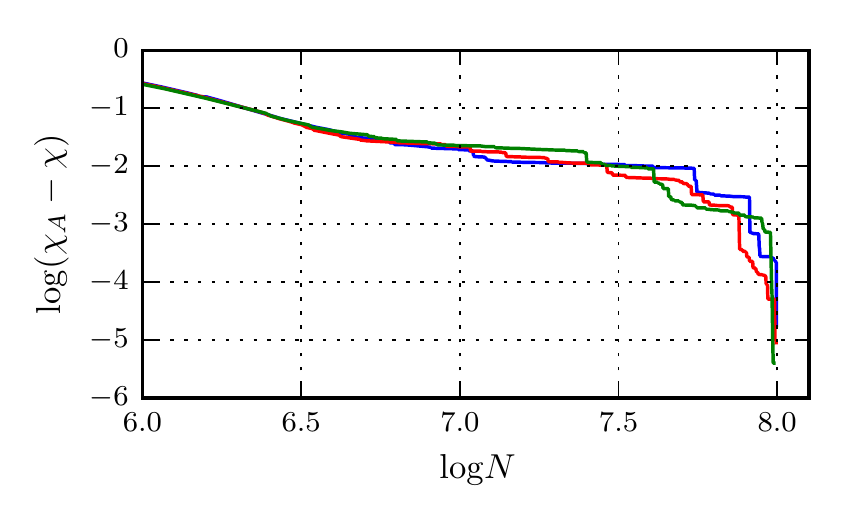}
  \caption{As in Fig. \ref{figlr13} (b) but now in log-log
plot to show that the approach to the asymptotic value $\chi_A$ is 
neither an exponential function nor a power law. The curves correspond to three
different initial conditions, $(s,p)= (4.0,0.75),\; (2.5,0.75),\; (2.5,0.15)$, in
green, blue and red correspondingly.}
  \label{figlr14}
\end{figure}

\noindent
It is clear that the random model is not good, and the approach to
the asymptotic value $\chi_A\approx 0.80$ is neither exponential nor
a power law, but something different to be studied further \cite{LR2017},
as one can see in the lin-lin plot (a) and in the log-lin plot in (b)
of Fig. \ref{figlr13}, and also in Fig. \ref{figlr14}  for
three different initial conditions. The selected initial conditions are well separated yet yield very similar results. We therefore expect an ensemble average would not change the overall shape of the curve.

One should note that the
measured $\chi (N)$ is always below the prediction of the random model,
which is due to the sticky objects in the phase space, that delay
the diffusion process, and occasionally also cause some plateaus on the
curve (due to temporary trapping). 
Also the cell occupancy numbers shown in (c) are different from
the simple binomial distribution for the cells. Clearly, the empty
cells at $M=0$ represent the regular part of the phase space, and all chaotic components not linked to the largest one. These cells remain permanently empty for all $N$.  There is an approximately Gaussian
distribution around the mean value $\mu = \langle M \rangle =N/N_c$, where
$N_c = \chi_A L^2$, with $L=1000$, and thus approximately $\mu \approx 62.5$,
for $L=1000$, $N=5\times 10^7$ and $\chi_A \approx 0.80$. The numerical
value of $\sigma^2$ is slightly larger, $\sigma^2=72$.
In between we observe a shallow minimum, sparsely populated. Further work along 
these lines is in progress \cite{LR2017}.

\section{Discussion and conclusions}

In conclusion we may say that the major aspects of global diffusion in the
stadium billiard are well understood, and that many aspects can be manifested 
in other systems with slow ergodicity. The applicability of the random model 
\cite{Rob1997} is
largely confirmed, the coarse grained phase space divided into cells
is being filled exponentially. The diffusion constant obeys the parabolic law
$D=D_0(\ep) (1-p^2)$, which we may expect to apply in other slow 
ergodic billiards as well. The distribution function 
emanating from an arbitrary initial
condition obeys very well the inhomogeneous diffusion equation, and the
diffusion is normal for all $\ep$ and initial conditions $p_0$. 
The boundary effects in the evolution of $\rho(p,t)$ 
are correctly described by the model. The approach to uniform  
equilibrium distribution $\rho=1/2$ with the variance $\Vp=1/3$ is  
always exponential, for all $\ep$ and all initial conditions $p_0$.
The diffusion constant $D_0$ has been calculated for many different values
of $\ep$. At small $\ep \le \ep_c \approx 0.1$ $D_0 \propto \ep^{5/2}$,
while for larger $\ep \ge \ep_c\approx 0.1$ it goes as $\propto \ep^2$, 
in agreement with the previous works \cite{BCL1996,CP1999}, but in between
there is no theoretical analytical approximation, so we have to resort to
the numerical calculations performed in this work. The value of the classical
transport (diffusion) time $N_T$, in terms of the discrete time (number
of collisions) has been determined for all values of $\ep\le0.2$, 
which plays an important role in
the quantum chaos of localized chaotic eigenstates 
\cite{BR2013,BR2013A,BLR2017}.

In the mixed type systems, exemplified by the billiard introduced in
the Ref. \cite{Rob1983}, with the shape parameter $\lambda=0.15$, we have 
shown that the behavior is quite different from ergodic fully chaotic 
systems, which is in agreement with the report of Meiss \cite{Mei1994}
on the standard map.

Further work along these lines is important for the understanding of
classical and quantum chaos in billiard systems as model systems, 
but the approach
should also be applicable to other smooth Hamiltonian systems, such
as e.g. the hydrogen atom in a strong magnetic field 
\cite{Rob1981,Rob1982,HRW1989,WF1989}, or the helium atom etc.

\section{Acknowledgement}

The authors acknowledge the financial support of the Slovenian Research Agency (research core funding P1-0306). We would like to thank Dr. Benjamin Batisti\'c for useful discussions an providing the use of his excellent numerical library available at {\em  https://github.com/benokit/time-dep-billiards}.

\bibliography{lozrob1}

\providecommand{\noopsort}[1]{}\providecommand{\singleletter}[1]{#1}%
\begin{thebibliography}{40}%
\makeatletter
\providecommand \@ifxundefined [1]{%
 \@ifx{#1\undefined}
}%
\providecommand \@ifnum [1]{%
 \ifnum #1\expandafter \@firstoftwo
 \else \expandafter \@secondoftwo
 \fi
}%
\providecommand \@ifx [1]{%
 \ifx #1\expandafter \@firstoftwo
 \else \expandafter \@secondoftwo
 \fi
}%
\providecommand \natexlab [1]{#1}%
\providecommand \enquote  [1]{``#1''}%
\providecommand \bibnamefont  [1]{#1}%
\providecommand \bibfnamefont [1]{#1}%
\providecommand \citenamefont [1]{#1}%
\providecommand \href@noop [0]{\@secondoftwo}%
\providecommand \href [0]{\begingroup \@sanitize@url \@href}%
\providecommand \@href[1]{\@@startlink{#1}\@@href}%
\providecommand \@@href[1]{\endgroup#1\@@endlink}%
\providecommand \@sanitize@url [0]{\catcode `\\12\catcode `\$12\catcode
  `\&12\catcode `\#12\catcode `\^12\catcode `\_12\catcode `\%12\relax}%
\providecommand \@@startlink[1]{}%
\providecommand \@@endlink[0]{}%
\providecommand \url  [0]{\begingroup\@sanitize@url \@url }%
\providecommand \@url [1]{\endgroup\@href {#1}{\urlprefix }}%
\providecommand \urlprefix  [0]{URL }%
\providecommand \Eprint [0]{\href }%
\providecommand \doibase [0]{http://dx.doi.org/}%
\providecommand \selectlanguage [0]{\@gobble}%
\providecommand \bibinfo  [0]{\@secondoftwo}%
\providecommand \bibfield  [0]{\@secondoftwo}%
\providecommand \translation [1]{[#1]}%
\providecommand \BibitemOpen [0]{}%
\providecommand \bibitemStop [0]{}%
\providecommand \bibitemNoStop [0]{.\EOS\space}%
\providecommand \EOS [0]{\spacefactor3000\relax}%
\providecommand \BibitemShut  [1]{\csname bibitem#1\endcsname}%
\let\auto@bib@innerbib\@empty
%</preamble>
\bibitem [{\citenamefont {Bunimovich}(1979)}]{Bun1979}%
  \BibitemOpen
  \bibfield  {author} {\bibinfo {author} {\bibfnamefont {L.~A.}\ \bibnamefont
  {Bunimovich}},\ }\href@noop {} {\bibfield  {journal} {\bibinfo  {journal}
  {Comm. Math. Phys.}\ }\textbf {\bibinfo {volume} {65}},\ \bibinfo {pages}
  {295} (\bibinfo {year} {1979})}\BibitemShut {NoStop}%
\bibitem [{\citenamefont {Berry}(1981)}]{Ber1981}%
  \BibitemOpen
  \bibfield  {author} {\bibinfo {author} {\bibfnamefont {M.~V.}\ \bibnamefont
  {Berry}},\ }\href@noop {} {\bibfield  {journal} {\bibinfo  {journal} {Eur. J.
  Phys.}\ }\textbf {\bibinfo {volume} {2}},\ \bibinfo {pages} {91} (\bibinfo
  {year} {1981})}\BibitemShut {NoStop}%
\bibitem [{\citenamefont {Benettin}(1984)}]{Ben1984}%
  \BibitemOpen
  \bibfield  {author} {\bibinfo {author} {\bibfnamefont {G.}~\bibnamefont
  {Benettin}},\ }\href@noop {} {\bibfield  {journal} {\bibinfo  {journal}
  {Physica}\ }\textbf {\bibinfo {volume} {13D}},\ \bibinfo {pages} {211}
  (\bibinfo {year} {1984})}\BibitemShut {NoStop}%
\bibitem [{\citenamefont {Borgonovi}\ \emph {et~al.}(1996)\citenamefont
  {Borgonovi}, \citenamefont {Casati},\ and\ \citenamefont {Li}}]{BCL1996}%
  \BibitemOpen
  \bibfield  {author} {\bibinfo {author} {\bibfnamefont {F.}~\bibnamefont
  {Borgonovi}}, \bibinfo {author} {\bibfnamefont {G.}~\bibnamefont {Casati}}, \
  and\ \bibinfo {author} {\bibfnamefont {B.}~\bibnamefont {Li}},\ }\href@noop
  {} {\bibfield  {journal} {\bibinfo  {journal} {Phys. Rev. Lett.}\ }\textbf
  {\bibinfo {volume} {77}},\ \bibinfo {pages} {4744} (\bibinfo {year}
  {1996})}\BibitemShut {NoStop}%
\bibitem [{\citenamefont {St\"ockmann}(1999)}]{Stoe}%
  \BibitemOpen
  \bibfield  {author} {\bibinfo {author} {\bibfnamefont {H.-J.}\ \bibnamefont
  {St\"ockmann}},\ }\href@noop {} {\emph {\bibinfo {title} {Quantum Chaos - An
  Introduction}}}\ (\bibinfo  {publisher} {Cambridge: Cambridge University
  Press},\ \bibinfo {year} {1999})\BibitemShut {NoStop}%
\bibitem [{\citenamefont {Haake}(2001)}]{Haake}%
  \BibitemOpen
  \bibfield  {author} {\bibinfo {author} {\bibfnamefont {F.}~\bibnamefont
  {Haake}},\ }\href@noop {} {\emph {\bibinfo {title} {Quantum Signatures of
  Chaos}}}\ (\bibinfo  {publisher} {Berlin: Springer},\ \bibinfo {year}
  {2001})\BibitemShut {NoStop}%
\bibitem [{\citenamefont {Robnik}(1998)}]{Rob1998}%
  \BibitemOpen
  \bibfield  {author} {\bibinfo {author} {\bibfnamefont {M.}~\bibnamefont
  {Robnik}},\ }\href@noop {} {\bibfield  {journal} {\bibinfo  {journal}
  {Nonlinear Phenomena in Complex Systems (Minsk)}\ }\textbf {\bibinfo {volume}
  {1}},\ \bibinfo {pages} {1} (\bibinfo {year} {1998})}\BibitemShut {NoStop}%
\bibitem [{\citenamefont {Batisti\'c}\ and\ \citenamefont
  {Robnik}(2010)}]{BR2010}%
  \BibitemOpen
  \bibfield  {author} {\bibinfo {author} {\bibfnamefont {B.}~\bibnamefont
  {Batisti\'c}}\ and\ \bibinfo {author} {\bibfnamefont {M.}~\bibnamefont
  {Robnik}},\ }\href@noop {} {\bibfield  {journal} {\bibinfo  {journal} {J.
  Phys. A: Math. Theor.}\ }\textbf {\bibinfo {volume} {43}},\ \bibinfo {pages}
  {215101} (\bibinfo {year} {2010})}\BibitemShut {NoStop}%
\bibitem [{\citenamefont {Batisti\'c}\ and\ \citenamefont
  {Robnik}(2013{\natexlab{a}})}]{BR2013}%
  \BibitemOpen
  \bibfield  {author} {\bibinfo {author} {\bibfnamefont {B.}~\bibnamefont
  {Batisti\'c}}\ and\ \bibinfo {author} {\bibfnamefont {M.}~\bibnamefont
  {Robnik}},\ }\href@noop {} {\bibfield  {journal} {\bibinfo  {journal} {J.
  Phys. A: Math. Theor.}\ }\textbf {\bibinfo {volume} {46}},\ \bibinfo {pages}
  {315102} (\bibinfo {year} {2013}{\natexlab{a}})}\BibitemShut {NoStop}%
\bibitem [{\citenamefont {Batisti\'c}\ and\ \citenamefont
  {Robnik}(2013{\natexlab{b}})}]{BR2013A}%
  \BibitemOpen
  \bibfield  {author} {\bibinfo {author} {\bibfnamefont {B.}~\bibnamefont
  {Batisti\'c}}\ and\ \bibinfo {author} {\bibfnamefont {M.}~\bibnamefont
  {Robnik}},\ }\href@noop {} {\bibfield  {journal} {\bibinfo  {journal} {Phys.
  Rev. E}\ }\textbf {\bibinfo {volume} {88}},\ \bibinfo {pages} {052913}
  (\bibinfo {year} {2013}{\natexlab{b}})}\BibitemShut {NoStop}%
\bibitem [{\citenamefont {Casati}\ \emph {et~al.}(1979)\citenamefont {Casati},
  \citenamefont {Chirikov}, \citenamefont {Izrailev},\ and\ \citenamefont
  {Ford}}]{Cas1979}%
  \BibitemOpen
  \bibfield  {author} {\bibinfo {author} {\bibfnamefont {G.}~\bibnamefont
  {Casati}}, \bibinfo {author} {\bibfnamefont {B.~V.}\ \bibnamefont
  {Chirikov}}, \bibinfo {author} {\bibfnamefont {F.~M.}\ \bibnamefont
  {Izrailev}}, \ and\ \bibinfo {author} {\bibfnamefont {J.}~\bibnamefont
  {Ford}},\ }\href@noop {} {\bibfield  {journal} {\bibinfo  {journal} {Lecture
  Notes in Physics}\ }\textbf {\bibinfo {volume} {93}},\ \bibinfo {pages} {334}
  (\bibinfo {year} {1979})}\BibitemShut {NoStop}%
\bibitem [{\citenamefont {Chirikov}\ \emph {et~al.}(1981)\citenamefont
  {Chirikov}, \citenamefont {Izrailev},\ and\ \citenamefont
  {Shepelyansky}}]{Chi1981}%
  \BibitemOpen
  \bibfield  {author} {\bibinfo {author} {\bibfnamefont {B.~V.}\ \bibnamefont
  {Chirikov}}, \bibinfo {author} {\bibfnamefont {F.~M.}\ \bibnamefont
  {Izrailev}}, \ and\ \bibinfo {author} {\bibfnamefont {D.~L.}\ \bibnamefont
  {Shepelyansky}},\ }\href@noop {} {\bibfield  {journal} {\bibinfo  {journal}
  {Sov. Sci. Rev. C}\ }\textbf {\bibinfo {volume} {2}},\ \bibinfo {pages} {209}
  (\bibinfo {year} {1981})}\BibitemShut {NoStop}%
\bibitem [{\citenamefont {Chirikov}\ \emph {et~al.}(1988)\citenamefont
  {Chirikov}, \citenamefont {Izrailev},\ and\ \citenamefont
  {Shepelyansky}}]{Chi1988}%
  \BibitemOpen
  \bibfield  {author} {\bibinfo {author} {\bibfnamefont {B.~V.}\ \bibnamefont
  {Chirikov}}, \bibinfo {author} {\bibfnamefont {F.~M.}\ \bibnamefont
  {Izrailev}}, \ and\ \bibinfo {author} {\bibfnamefont {D.~L.}\ \bibnamefont
  {Shepelyansky}},\ }\href@noop {} {\bibfield  {journal} {\bibinfo  {journal}
  {Physica D}\ }\textbf {\bibinfo {volume} {33}},\ \bibinfo {pages} {77}
  (\bibinfo {year} {1988})}\BibitemShut {NoStop}%
\bibitem [{\citenamefont {Izrailev}(1990)}]{Izr1990}%
  \BibitemOpen
  \bibfield  {author} {\bibinfo {author} {\bibfnamefont {F.~M.}\ \bibnamefont
  {Izrailev}},\ }\href@noop {} {\bibfield  {journal} {\bibinfo  {journal}
  {Phys. Rep.}\ }\textbf {\bibinfo {volume} {196}},\ \bibinfo {pages} {299}
  (\bibinfo {year} {1990})}\BibitemShut {NoStop}%
\bibitem [{\citenamefont {Izrailev}(1988)}]{Izr1988}%
  \BibitemOpen
  \bibfield  {author} {\bibinfo {author} {\bibfnamefont {F.~M.}\ \bibnamefont
  {Izrailev}},\ }\href@noop {} {\bibfield  {journal} {\bibinfo  {journal}
  {Phys. Lett. A}\ }\textbf {\bibinfo {volume} {134}},\ \bibinfo {pages} {13}
  (\bibinfo {year} {1988})}\BibitemShut {NoStop}%
\bibitem [{\citenamefont {Izrailev}(1989)}]{Izr1989}%
  \BibitemOpen
  \bibfield  {author} {\bibinfo {author} {\bibfnamefont {F.~M.}\ \bibnamefont
  {Izrailev}},\ }\href@noop {} {\bibfield  {journal} {\bibinfo  {journal} {J.
  Phys. A: Math. Gen.}\ }\textbf {\bibinfo {volume} {22}},\ \bibinfo {pages}
  {865} (\bibinfo {year} {1989})}\BibitemShut {NoStop}%
\bibitem [{\citenamefont {Batisti{\'c}}\ \emph {et~al.}(2013)\citenamefont
  {Batisti{\'c}}, \citenamefont {Manos},\ and\ \citenamefont
  {Robnik}}]{BatManRob2013}%
  \BibitemOpen
  \bibfield  {author} {\bibinfo {author} {\bibfnamefont {B.}~\bibnamefont
  {Batisti{\'c}}}, \bibinfo {author} {\bibfnamefont {T.}~\bibnamefont {Manos}},
  \ and\ \bibinfo {author} {\bibfnamefont {M.}~\bibnamefont {Robnik}},\
  }\href@noop {} {\bibfield  {journal} {\bibinfo  {journal} {Europhys. Lett.}\
  }\textbf {\bibinfo {volume} {102}},\ \bibinfo {pages} {50008} (\bibinfo
  {year} {2013})}\BibitemShut {NoStop}%
\bibitem [{\citenamefont {Manos}\ and\ \citenamefont
  {Robnik}(2013)}]{ManRob2013}%
  \BibitemOpen
  \bibfield  {author} {\bibinfo {author} {\bibfnamefont {T.}~\bibnamefont
  {Manos}}\ and\ \bibinfo {author} {\bibfnamefont {M.}~\bibnamefont {Robnik}},\
  }\href@noop {} {\bibfield  {journal} {\bibinfo  {journal} {Phys. Rev. E}\
  }\textbf {\bibinfo {volume} {87}},\ \bibinfo {pages} {062905} (\bibinfo
  {year} {2013})}\BibitemShut {NoStop}%
\bibitem [{\citenamefont {Manos}\ and\ \citenamefont
  {Robnik}(2014)}]{ManRob2014}%
  \BibitemOpen
  \bibfield  {author} {\bibinfo {author} {\bibfnamefont {T.}~\bibnamefont
  {Manos}}\ and\ \bibinfo {author} {\bibfnamefont {M.}~\bibnamefont {Robnik}},\
  }\href@noop {} {\bibfield  {journal} {\bibinfo  {journal} {Phys. Rev. E}\
  }\textbf {\bibinfo {volume} {89}},\ \bibinfo {pages} {022905} (\bibinfo
  {year} {2014})}\BibitemShut {NoStop}%
\bibitem [{\citenamefont {Manos}\ and\ \citenamefont
  {Robnik}(2015)}]{ManRob2015}%
  \BibitemOpen
  \bibfield  {author} {\bibinfo {author} {\bibfnamefont {T.}~\bibnamefont
  {Manos}}\ and\ \bibinfo {author} {\bibfnamefont {M.}~\bibnamefont {Robnik}},\
  }\href@noop {} {\bibfield  {journal} {\bibinfo  {journal} {Phys. Rev. E}\
  }\textbf {\bibinfo {volume} {91}},\ \bibinfo {pages} {042904} (\bibinfo
  {year} {2015})}\BibitemShut {NoStop}%
\bibitem [{\citenamefont {Batisti\'c}\ \emph {et~al.}(2017)\citenamefont
  {Batisti\'c}, \citenamefont {\v{C}. Lozej},\ and\ \citenamefont
  {Robnik}}]{BLR2017}%
  \BibitemOpen
  \bibfield  {author} {\bibinfo {author} {\bibfnamefont {B.}~\bibnamefont
  {Batisti\'c}}, \bibinfo {author} {\bibnamefont {\v{C}. Lozej}}, \ and\
  \bibinfo {author} {\bibfnamefont {M.}~\bibnamefont {Robnik}},\ }\href@noop {}
  {\bibfield  {journal} {\bibinfo  {journal} {in preparation}\ } (\bibinfo
  {year} {2017})}\BibitemShut {NoStop}%
\bibitem [{\citenamefont {Casati}\ and\ \citenamefont {Prosen}(1999)}]{CP1999}%
  \BibitemOpen
  \bibfield  {author} {\bibinfo {author} {\bibfnamefont {G.}~\bibnamefont
  {Casati}}\ and\ \bibinfo {author} {\bibfnamefont {T.}~\bibnamefont
  {Prosen}},\ }\href@noop {} {\bibfield  {journal} {\bibinfo  {journal}
  {Physica D}\ }\textbf {\bibinfo {volume} {131}},\ \bibinfo {pages} {293}
  (\bibinfo {year} {1999})}\BibitemShut {NoStop}%
\bibitem [{\citenamefont {Prosen}(2000)}]{Pro2000}%
  \BibitemOpen
  \bibfield  {author} {\bibinfo {author} {\bibfnamefont {T.}~\bibnamefont
  {Prosen}},\ }\href@noop {} {\emph {\bibinfo {title} {in Proc. of the Int.
  School in Phys. "Enrico Fermi", Course CXLIII, Eds. G. Casati and U.
  Smilansky}}}\ (\bibinfo  {publisher} {Amsterdam: IOS Press},\ \bibinfo {year}
  {2000})\BibitemShut {NoStop}%
\bibitem [{\citenamefont {Dana}\ \emph {et~al.}(1989)\citenamefont {Dana},
  \citenamefont {Murray},\ and\ \citenamefont {Percival}}]{DMP1989}%
  \BibitemOpen
  \bibfield  {author} {\bibinfo {author} {\bibfnamefont {I.}~\bibnamefont
  {Dana}}, \bibinfo {author} {\bibfnamefont {N.~W.}\ \bibnamefont {Murray}}, \
  and\ \bibinfo {author} {\bibfnamefont {I.~C.}\ \bibnamefont {Percival}},\
  }\href@noop {} {\bibfield  {journal} {\bibinfo  {journal} {Phys. Rev. Lett.}\
  }\textbf {\bibinfo {volume} {62}},\ \bibinfo {pages} {233} (\bibinfo {year}
  {1989})}\BibitemShut {NoStop}%
\bibitem [{\citenamefont {Chen}\ \emph {et~al.}(1990)\citenamefont {Chen},
  \citenamefont {Dana}, \citenamefont {Meiss}, \citenamefont {Murray},\ and\
  \citenamefont {Percival}}]{CDMMP1990}%
  \BibitemOpen
  \bibfield  {author} {\bibinfo {author} {\bibfnamefont {Q.}~\bibnamefont
  {Chen}}, \bibinfo {author} {\bibfnamefont {I.}~\bibnamefont {Dana}}, \bibinfo
  {author} {\bibfnamefont {J.~D.}\ \bibnamefont {Meiss}}, \bibinfo {author}
  {\bibfnamefont {N.~W.}\ \bibnamefont {Murray}}, \ and\ \bibinfo {author}
  {\bibfnamefont {I.~C.}\ \bibnamefont {Percival}},\ }\href@noop {} {\bibfield
  {journal} {\bibinfo  {journal} {Physica D}\ }\textbf {\bibinfo {volume}
  {46}},\ \bibinfo {pages} {217} (\bibinfo {year} {1990})}\BibitemShut
  {NoStop}%
\bibitem [{\citenamefont {Meiss}(1994)}]{Mei1994}%
  \BibitemOpen
  \bibfield  {author} {\bibinfo {author} {\bibfnamefont {J.~D.}\ \bibnamefont
  {Meiss}},\ }\href@noop {} {\bibfield  {journal} {\bibinfo  {journal} {Physica
  D}\ }\textbf {\bibinfo {volume} {74}},\ \bibinfo {pages} {254} (\bibinfo
  {year} {1994})}\BibitemShut {NoStop}%
\bibitem [{\citenamefont {Robnik}(1983)}]{Rob1983}%
  \BibitemOpen
  \bibfield  {author} {\bibinfo {author} {\bibfnamefont {M.}~\bibnamefont
  {Robnik}},\ }\href@noop {} {\bibfield  {journal} {\bibinfo  {journal} {J.
  Phys. A: Math. Gen.}\ }\textbf {\bibinfo {volume} {16}},\ \bibinfo {pages}
  {3971} (\bibinfo {year} {1983})}\BibitemShut {NoStop}%
\bibitem [{\citenamefont {Polyanin}(2002)}]{Pol2002}%
  \BibitemOpen
  \bibfield  {author} {\bibinfo {author} {\bibfnamefont {A.~D.}\ \bibnamefont
  {Polyanin}},\ }\href@noop {} {\emph {\bibinfo {title} {Handbook of linear
  partial differential equations for engineers and scientists}}}\ (\bibinfo
  {publisher} {Boca Raton: Chapman \& Hall/CRC},\ \bibinfo {year}
  {2002})\BibitemShut {NoStop}%
\bibitem [{\citenamefont {Lau}\ and\ \citenamefont {Lubensky}(2007)}]{LL2007}%
  \BibitemOpen
  \bibfield  {author} {\bibinfo {author} {\bibfnamefont {A.~V.~C.}\
  \bibnamefont {Lau}}\ and\ \bibinfo {author} {\bibfnamefont {T.~C.}\
  \bibnamefont {Lubensky}},\ }\href@noop {} {\bibfield  {journal} {\bibinfo
  {journal} {Phys. Rev. E}\ }\textbf {\bibinfo {volume} {76}},\ \bibinfo
  {pages} {011123} (\bibinfo {year} {2007})}\BibitemShut {NoStop}%
\bibitem [{\citenamefont {Santal\'o}\ and\ \citenamefont
  {Kac}(2004)}]{Santalo}%
  \BibitemOpen
  \bibfield  {author} {\bibinfo {author} {\bibfnamefont {L.~A.}\ \bibnamefont
  {Santal\'o}}\ and\ \bibinfo {author} {\bibfnamefont {M.}~\bibnamefont
  {Kac}},\ }\href@noop {} {\emph {\bibinfo {title} {Integral geometry and
  geometric probability}}},\ Cambridge mathematical library\ (\bibinfo
  {publisher} {Cambridge: Cambridge University Press},\ \bibinfo {year}
  {2004})\BibitemShut {NoStop}%
\bibitem [{\citenamefont {Vivaldi}\ \emph {et~al.}(1983)\citenamefont
  {Vivaldi}, \citenamefont {Casati},\ and\ \citenamefont {Guarneri}}]{VCG1983}%
  \BibitemOpen
  \bibfield  {author} {\bibinfo {author} {\bibfnamefont {F.}~\bibnamefont
  {Vivaldi}}, \bibinfo {author} {\bibfnamefont {G.}~\bibnamefont {Casati}}, \
  and\ \bibinfo {author} {\bibfnamefont {I.}~\bibnamefont {Guarneri}},\
  }\href@noop {} {\bibfield  {journal} {\bibinfo  {journal} {Phys. Rev. Lett.}\
  }\textbf {\bibinfo {volume} {51}},\ \bibinfo {pages} {727} (\bibinfo {year}
  {1983})}\BibitemShut {NoStop}%
\bibitem [{\citenamefont {Armstead}\ \emph {et~al.}(2004)\citenamefont
  {Armstead}, \citenamefont {Hunt},\ and\ \citenamefont {Ott}}]{AHO2004}%
  \BibitemOpen
  \bibfield  {author} {\bibinfo {author} {\bibfnamefont {D.~N.}\ \bibnamefont
  {Armstead}}, \bibinfo {author} {\bibfnamefont {B.}~\bibnamefont {Hunt}}, \
  and\ \bibinfo {author} {\bibfnamefont {E.}~\bibnamefont {Ott}},\ }\href@noop
  {} {\bibfield  {journal} {\bibinfo  {journal} {Physica D}\ }\textbf {\bibinfo
  {volume} {193}},\ \bibinfo {pages} {96} (\bibinfo {year} {2004})}\BibitemShut
  {NoStop}%
\bibitem [{\citenamefont {Altmann}\ \emph {et~al.}(2008)\citenamefont
  {Altmann}, \citenamefont {Friedrich}, \citenamefont {Motter}, \citenamefont
  {Kantz},\ and\ \citenamefont {Richter}}]{Alt2008}%
  \BibitemOpen
  \bibfield  {author} {\bibinfo {author} {\bibfnamefont {E.~G.}\ \bibnamefont
  {Altmann}}, \bibinfo {author} {\bibfnamefont {T.}~\bibnamefont {Friedrich}},
  \bibinfo {author} {\bibfnamefont {A.~E.}\ \bibnamefont {Motter}}, \bibinfo
  {author} {\bibfnamefont {H.}~\bibnamefont {Kantz}}, \ and\ \bibinfo {author}
  {\bibfnamefont {A.}~\bibnamefont {Richter}},\ }\href@noop {} {\bibfield
  {journal} {\bibinfo  {journal} {Phys. Rev. E}\ }\textbf {\bibinfo {volume}
  {77}},\ \bibinfo {pages} {016205} (\bibinfo {year} {2008})}\BibitemShut
  {NoStop}%
\bibitem [{\citenamefont {Markarian}(1993)}]{Mar1993}%
  \BibitemOpen
  \bibfield  {author} {\bibinfo {author} {\bibfnamefont {R.}~\bibnamefont
  {Markarian}},\ }\href@noop {} {\bibfield  {journal} {\bibinfo  {journal}
  {Nonlinearity}\ }\textbf {\bibinfo {volume} {6}},\ \bibinfo {pages} {819}
  (\bibinfo {year} {1993})}\BibitemShut {NoStop}%
\bibitem [{\citenamefont {Robnik}\ \emph {et~al.}(1997)\citenamefont {Robnik},
  \citenamefont {Dobnikar}, \citenamefont {Rapisarda}, \citenamefont {Prosen},\
  and\ \citenamefont {Petkov\v{s}ek}}]{Rob1997}%
  \BibitemOpen
  \bibfield  {author} {\bibinfo {author} {\bibfnamefont {M.}~\bibnamefont
  {Robnik}}, \bibinfo {author} {\bibfnamefont {J.}~\bibnamefont {Dobnikar}},
  \bibinfo {author} {\bibfnamefont {A.}~\bibnamefont {Rapisarda}}, \bibinfo
  {author} {\bibfnamefont {T.}~\bibnamefont {Prosen}}, \ and\ \bibinfo {author}
  {\bibfnamefont {M.}~\bibnamefont {Petkov\v{s}ek}},\ }\href@noop {} {\bibfield
   {journal} {\bibinfo  {journal} {J. Phys. A: Math. Gen.}\ }\textbf {\bibinfo
  {volume} {30}},\ \bibinfo {pages} {L803} (\bibinfo {year}
  {1997})}\BibitemShut {NoStop}%
\bibitem [{\citenamefont {\v{C}. Lozej}\ and\ \citenamefont
  {Robnik}(2017)}]{LR2017}%
  \BibitemOpen
  \bibfield  {author} {\bibinfo {author} {\bibnamefont {\v{C}. Lozej}}\ and\
  \bibinfo {author} {\bibfnamefont {M.}~\bibnamefont {Robnik}},\ }\href@noop {}
  {\bibfield  {journal} {\bibinfo  {journal} {in preparation}\ } (\bibinfo
  {year} {2017})}\BibitemShut {NoStop}%
\bibitem [{\citenamefont {Robnik}(1981)}]{Rob1981}%
  \BibitemOpen
  \bibfield  {author} {\bibinfo {author} {\bibfnamefont {M.}~\bibnamefont
  {Robnik}},\ }\href@noop {} {\bibfield  {journal} {\bibinfo  {journal} {J.
  Phys. A: Math. Gen.}\ }\textbf {\bibinfo {volume} {14}},\ \bibinfo {pages}
  {3195} (\bibinfo {year} {1981})}\BibitemShut {NoStop}%
\bibitem [{\citenamefont {Robnik}(1982)}]{Rob1982}%
  \BibitemOpen
  \bibfield  {author} {\bibinfo {author} {\bibfnamefont {M.}~\bibnamefont
  {Robnik}},\ }\href@noop {} {\bibfield  {journal} {\bibinfo  {journal} {J.
  Phys. Colloque C2}\ }\textbf {\bibinfo {volume} {43}},\ \bibinfo {pages} {29}
  (\bibinfo {year} {1982})}\BibitemShut {NoStop}%
\bibitem [{\citenamefont {Hasegawa}\ \emph {et~al.}(1989)\citenamefont
  {Hasegawa}, \citenamefont {Robnik},\ and\ \citenamefont {Wunner}}]{HRW1989}%
  \BibitemOpen
  \bibfield  {author} {\bibinfo {author} {\bibfnamefont {H.}~\bibnamefont
  {Hasegawa}}, \bibinfo {author} {\bibfnamefont {M.}~\bibnamefont {Robnik}}, \
  and\ \bibinfo {author} {\bibfnamefont {G.}~\bibnamefont {Wunner}},\
  }\href@noop {} {\bibfield  {journal} {\bibinfo  {journal} {Prog. Theor. Phys.
  Suppl. (Kyoto)}\ }\textbf {\bibinfo {volume} {98}},\ \bibinfo {pages} {198}
  (\bibinfo {year} {1989})}\BibitemShut {NoStop}%
\bibitem [{\citenamefont {Wintgen}\ and\ \citenamefont
  {Friedrich}(1989)}]{WF1989}%
  \BibitemOpen
  \bibfield  {author} {\bibinfo {author} {\bibfnamefont {D.}~\bibnamefont
  {Wintgen}}\ and\ \bibinfo {author} {\bibfnamefont {H.}~\bibnamefont
  {Friedrich}},\ }\href@noop {} {\bibfield  {journal} {\bibinfo  {journal}
  {Phys. Rep.}\ }\textbf {\bibinfo {volume} {183}},\ \bibinfo {pages} {38}
  (\bibinfo {year} {1989})}\BibitemShut {NoStop}%
\end{thebibliography}%

\end{document}